\documentclass[12pt,preprint]{emulateapj}

\shorttitle{Accretion and AGN Unification}
\shortauthors{Trump et al.}

\begin{document}

%\slugcomment{\bf Draft: \today}

\title{Accretion Rate and the Physical Nature of Unobscured Active
  Galaxies\altaffilmark{1}}

\author{
  Jonathan R. Trump,\altaffilmark{2}
  Christopher D. Impey,\altaffilmark{2}
%%  Martin Elvis,\altaffilmark{3}
  Brandon C. Kelly,\altaffilmark{3}$^,$\altaffilmark{4}
  Francesca Civano,\altaffilmark{3}
  Jared M. Gabor,\altaffilmark{2}
  Aleksandar M. Diamond-Stanic,\altaffilmark{2}
  Andrea Merloni,\altaffilmark{5}
  C. Megan Urry,\altaffilmark{6}
  Heng Hao,\altaffilmark{3}
  Knud Jahnke,\altaffilmark{7}
  Tohru Nagao,\altaffilmark{8}
  Yoshi Taniguchi,\altaffilmark{8}
%%  Marcella Brusa,\altaffilmark{6}
  Anton M. Koekemoer,\altaffilmark{9}
  Giorgio Lanzuisi,\altaffilmark{3}
  Charles Liu,\altaffilmark{10}$^,$\altaffilmark{11}
  Vincenzo Mainieri,\altaffilmark{12}
  Mara Salvato,\altaffilmark{5}
  and Nick Z. Scoville\altaffilmark{13}
}

\altaffiltext{1}{
  Based on observations with the XMM-Newton Observatory, an ESA
  science mission with instruments and contributions directly funded
  by ESA Member States and NASA; the Magellan Telescope, operated by
  the Carnegie Observatories; the European Southern Observatory (ESO)
  Very Large Telescope (VLT); and the MMT Observatory, a joint
  facility of the University of Arizona and the Smithsonian
  Institution; the Subaru Telescope, operated by the National
  Astronomical Observatory of Japan; and the NASA/ESA \emph{Hubble
  Space Telescope}, operated at the Space Telescope Science
  Institute, which is operated by AURA Inc, under NASA contract NAS
  5-26555.
\label{cosmos}}

\altaffiltext{2}{
  Steward Observatory, University of Arizona, 933 North Cherry Avenue,
  Tucson, AZ 85721 USA
\label{Arizona}}

\altaffiltext{3}{
  Harvard-Smithsonian Center for Astrophysics, 60 Garden Street,
  Cambridge, MA 02138 USA
\label{CfA}}

\altaffiltext{4}{
  Hubble Fellow
\label{Hubble}}

\altaffiltext{5}{
  Max Planck-Institut f\"ur Extraterrestrische Physik,
  Giessenbachstrasse 1, D-85748 Garching, Germany
\label{Max Planck}}

\altaffiltext{6}{
  Physics Department and Yale Center for Astronomy and Astrophysics,
  Yale University, New Haven, CT 06511, USA
\label{Yale}}

\altaffiltext{7}{
  Max Planck Institut f\"ur Astronomie, K\"onigstuhl 17, D-69117
  Heidelberg, Germany
\label{Max Planck 2}}

\altaffiltext{8}{
  Research Center for Space and Cosmic Evolution, Ehime University,
  2-5 Bunkyo-cho, Matsuyama 790-8577, Japan
\label{Ehime}}

\altaffiltext{9}{
  Space Telescope Science Institute, 3700 San Martin Drive, Baltimore,
  MD 21218 USA
\label{STScI}}

\altaffiltext{10}{
  Astrophysical Observatory, Department of Engineering Science and
  Physics, City University of New York, College of Staten Island, 2800
  Victory Blvd., Staten Island, NY 10314
\label{CUNY}}

\altaffiltext{11}{
  Hayden Planetarium, American Museum of Natural History, Central Park
  West at 79th Street, New York, NY 10024 USA
\label{AMNH}}

\altaffiltext{12}{
  European Southern Observatory, Karl-Schwarschild-Strasse 2, D-85748
  Garching, Germany
\label{ESO}}

\altaffiltext{13}{
  California Institute of Technology, MC 105-24, 1200 East California
  Boulevard, Pasadena, CA 91125 USA
\label{Caltech}}

%% \altaffiltext{6}{
%%   University of Maryland, Baltimore County, 1000 Hilltop Circle,
%%   Baltimore, MD 21250
%% \label{Maryland}}
%% 
%% \altaffiltext{7}{
%%   Observatories of the Carnegie Institute of Washington, Santa Barbara
%%   Street, Pasadena, CA 91101
%% \label{Carnegie}}

\def\etal{et al.}
\newcommand{\Ha}{\hbox{{\rm H}$\alpha$}}
\newcommand{\Hb}{\hbox{{\rm H}$\beta$}}
\newcommand{\MgII}{\hbox{{\rm Mg}\kern 0.1em{\sc ii}}}
\newcommand{\CIII}{\hbox{{\rm C}\kern 0.1em{\sc iii}]}}
\newcommand{\CIV}{\hbox{{\rm C}\kern 0.1em{\sc iv}}}
\newcommand{\OII}{\hbox{[{\rm O}\kern 0.1em{\sc ii}]}}
\newcommand{\OIII}{\hbox{[{\rm O}\kern 0.1em{\sc iii}]}}
\newcommand{\OIV}{\hbox{[{\rm O}\kern 0.1em{\sc iv}]}}
\newcommand{\NII}{\hbox{[{\rm N}\kern 0.1em{\sc ii}]}}

\begin{abstract}
  We show how accretion rate governs the physical properties of a
  sample of unobscured broad-line, narrow-line, and lineless active
  galactic nuclei (AGNs).  We avoid the systematic errors plaguing
  previous studies of AGN accretion rate by using accurate accretion
  luminosities ($L_{int}$) from well-sampled multiwavelength SEDs from
  the Cosmic Evolution Survey (COSMOS), and accurate black hole masses
  derived from virial scaling relations (for broad-line AGNs) or
  host-AGN relations (for narrow-line and lineless AGNs).  In general,
  broad emission lines are present only at the highest accretion rates
  ($L_{int}/L_{Edd} > 10^{-2}$), and these rapidly accreting AGNs are
  observed as broad-line AGNs or possibly as obscured narrow-line
  AGNs.  Narrow-line and lineless AGNs at lower specific accretion
  rates ($L_{int}/L_{Edd} < 10^{-2}$) are unobscured and yet lack a
  broad line region.  The disappearance of the broad emission lines is
  caused by an expanding radiatively inefficient accretion flow (RIAF)
  at the inner radius of the accretion disk.  The presence of the RIAF
  also drives $L_{int}/L_{Edd} < 10^{-2}$ narrow-line and lineless
  AGNs to 10 times higher ratios of radio to optical/UV emission than
  $L_{int}/L_{Edd} > 10^{-2}$ broad-line AGNs, since the unbound
  nature of the RIAF means it is easier to form a radio outflow.  The
  IR torus signature also tends to become weaker or disappear from
  $L_{int}/L_{Edd} < 10^{-2}$ AGNs, although there may be additional
  mid-IR synchrotron emission associated with the RIAF.  Together
  these results suggest that specific accretion rate is an important
  physical ``axis'' of AGN unification, described by a simple model.
\end{abstract}


\section{Introduction}

Supermassive black holes (SMBHs) are now known to be ubiquitous in the
centers of all massive galaxies \citep{mag98}.  SMBHs grow in an
``active'' phase of accretion, during which they are observed as
active galactic nuclei (AGN).  AGN growth is intimately tied to galaxy
evolution, as evident in the well-studied correlations between SMBH
mass ($M_{BH}$) and properties of the host galaxy bulge
\citep[e.g.,][]{geb00, fer00, mar03}.  The AGN phase is also
hypothesized to regulate star formation in its host galaxy, with the
galaxy feeding the black hole in turn \citep[e.g.][]{dim05, you08}.
All massive galaxies are thought to experience episodic AGN behavior
in their lifetime \citep{sol82, mar04}.

AGNs are generally classified by differences in their optical spectra.
Type 1 or broad-line AGNs have broad ($v_{FWHM} \gtrsim
1000$~km~s$^{-1}$) emission lines superimposed on blue unobscured
continua in the UV/optical \citep[e.g.,][]{van01}, and are the most
luminous persistent sources in the sky.  Type 2 or narrow-line AGNs
lack broad emission lines and have weaker continua (frequently
dominated by their host galaxies), but have strong narrow emission
lines, especially from forbidden transitions.  Narrow emission lines
associated with nuclear activity can be distinguished from lines
caused by star formation by studying the line ratios \citep{bpt81}.
The line ratio diagnostics work because the ``harder'' emission of an
AGN is more efficient at ionizing the surrounding gas and dust than
star formation, and thus AGNs have stronger lines from high-energy
forbidden transitions (e.g., \OIII~$\lambda$5007\AA~and
\NII~$\lambda$6583\AA) relative to the lower-energy hydrogen
transitions (e.g., \Hb~$\lambda$4861\AA~and \Ha~$\lambda$6563\AA).
The subclass of ``low-ionization nuclear emission region'' AGNs
\citep[LINERs,][]{heck80} have narrow emission lines that are probably
excited by some combination of ionization from both star formation and
an AGN \citep{era10}.  Deep X-ray surveys have additionally revealed
``optically dull'' AGNs \citep{elv81,com02}, which have bright X-ray
emission but none of the broad or narrow emission line signatures of
AGN accretion.  While many optically dull AGNs can be explained as
Type 2 AGNs diluted by prominent host galaxies \citep{mor02, cac07},
at least $\sim$1/3 are undiluted but intrinsically optically weaker
than other AGNs \citep{tru09c}.  The inferred X-ray column density
$N_H$ can also be used to classify AGNs, with Type 2 (narrow-line)
AGNs typically more X-ray absorbed than Type 1 (broad-line) AGNs.
However X-ray and optical classifications differ for $\sim$20\% of
objects \citep{tro09}.

Historically, Type 2 and optically dull AGNs have been described as
obscured versions of Type 1 AGN, with the broad emission line region
(BLR) hidden behind a partially opaque ``torus'' of gas and dust,
while the narrow emission lines lie outside the torus
\citep[e.g.,][]{kro88}.  The best evidence for this scenario is the
observation that some Type 2 AGNs have a ``hidden'' BLR revealed by
spectropolarimetry \citep{ant93}.  However, recent observations have
revealed several serious limitations of a simple unified model based
solely on geometric obscuration.  Even in very deep
spectropolarimetric observations, many Type 2 AGNs show no hidden BLR
\citep{barth99,tran01,wang07}.  Observations suggest a lower
$L/L_{Edd} \ge 0.01$ limit in accretion rate for broad-line AGNs
\citep{kol06,tru09b}, although they remain incomplete at low accretion
rates and low masses \citep{kel10}.  The X-ray spectra are unabsorbed
($N_H \lesssim 10^{21}$~cm$^{-2}$) for 30-40\% of Type 2 AGNs
\citep{main07,tro09}, as well as most local LINERs \citep[][and
references therein]{ho08} and distant optically dull AGNs
\citep{tru09c}.  Several well-studied LINERs additionally lack the
narrow Fe K$\alpha$ emission signature of a dusty torus
\citep{ptak04,bia08}.  Many Type 2 AGNs and most optically dull AGNs
have mid-IR colors like normal galaxies \citep{ho08,tru09c}, in
contrast to hot mid-IR colors of Type 1 AGNs described by torus models
\citep{nen08,mor09}.  Toroidal obscuration is additionally ruled out
for some strongly varying Type 2 \citep{haw04} and optically dull AGNs
\citep{tru09c}, since these objects have continua which vary on year
timescales, well within the inferred light travel time dimension of
any torus.

Several authors have proposed models which use different accretion
rates as a cause of the differences between observed AGNs.
\citet{eli09} suggest that the BLR and ``torus'' are inner (ionized)
and outer (clumpy and dusty) parts of the same disk-driven wind, and
that this wind is no longer supported at low accretion rate \citep[see
  also][]{eli06,nen08}.  Similarly, \citet{nic00} suggested that low
accretion rates actually drive the disk wind within the last stable
orbit of the SMBH, meaning that the BLR cannot form.  Models for
radiatively inefficient accretion \citep[e.g.,][]{yuan07} suggest that
at $L/L_{Edd} \lesssim 10^{-2}$, the accretion disk becomes truncated
near the SMBH, with a geometrically thick and optically thin disk at
low radii, and a normal thin disk \citep[e.g.,][]{sha73} at higher
radii.  Such objects are predicted to lack strong emission lines (both
broad and narrow) and have weak UV/optical emission, as observed in
many optically weak low-luminosity AGNs \citep{ho09} and X-ray bright,
optically dull AGNs \citep{tru09c}.  \citet{hop09} additionally show
that X-ray hardness, generally attributed to X-ray absorption, may
also result from the naturally X-ray hard spectrum expected from
radiatively inefficient accretion.

In this work we directly measure Eddington ratios for a large, X-ray
selected sample of broad-line, narrow-line, and lineless AGNs.  The
Eddington ratio is a unitless measure of accretion power, defined as
$\lambda \equiv L_{int}/L_{Edd}$. (with $L_{int}$ the intrinsic
accretion luminosity).  The sample is drawn from the Cosmic Evolution
Survey \citep[COSMOS,][]{sco07} X-ray AGN sample \citep{tru07}, as
described in Section 2.  Estimates of specific accretion rates are
described in Section 3, with intrinsic accretion luminosity $L_{int}$
measured directly from fits to the multiwavelength continuum (avoiding
uncertain bolometric corrections) and black hole masses from the broad
line scaling relations (for broad-line AGNs) or the $M_{BH}-M_*$
relations (for narrow-line and lineless AGNs).  In Section 4 we show
that broad emission lines are present at only high accretion rates
($L_{int}/L_{Edd} > 0.01$), while narrow-line and lineless AGNs at
lower accretion rates have cooler disks, stronger radio jets, and no
torus IR signature.  We present a ``cartoon'' model which summarizes
our results in Section 5, with predictions for future observations in
Section 6.  We adopt a cosmology with $h=0.70$, $\Omega_M=0.3$,
$\Omega_{\Lambda}=0.7$ throughout.

\section{Observational Data}

Measuring an accurate specific accretion rate requires accurate
accretion luminosities and black hole mass estimates.  In particular,
SED measurements from optical/UV to X-ray are necessary to constrain
intrinsic luminosities to within a factor of a few (as we show in \S
3.1).  We select a sample of 348 AGNs from the Cosmic Evolution Survey
\citep[COSMOS,][]{sco07} field, which is based on the 1.7 deg$^2$
HST/ACS mosaic \citep{koe07}.  These AGNs have multiwavelength data in
the form of Spitzer/IRAC, HST/ACS, Subaru/Suprime-Cam, GALEX,
XMM-Newton, and Chandra observations, as described in Table
\ref{tbl:obsdata}.  Spectroscopic identification and redshifts for
these objects comes from archival SDSS data, Magellan/IMACS and
MMT/Hectospec \citep{tru09a}, and VLT/VIMOS observations
\citep{lil10}.

The sample is selected from the parent catalog of 1651 XMM-COSMOS
point sources with optical counterparts \citep{bru10}, limited by
$f_{0.5-2 {\rm keV}} > 2 \times 10^{-16}$ erg~s$^{-1}$~cm$^{-2}$.  Of
these X-ray point sources, 649 objects with $i_{\rm AB}<23.5$ have
high-confidence ($>90\%$ likelihood as correct) identifications and
redshifts from optical spectroscopy \citep{tru09a,lil10} in COSMOS.
Most of the X-ray point sources without spectroscopy were missed
simply due to random slit placement constraints.  The optical
spectroscopy is $\sim$90\% complete to $i_{\rm AB}<22.5$, although the
completeness is redshift-dependent.  For broad-line AGNs, the
spectroscopic completeness is lower at $0.5<z<1$, $z \sim 1.4$, and $z
\sim 2.4$, especially at $i_{\rm AB}>22.5$ \citep[see Figure 13
of][]{tru09a}.  For narrow-line and lineless AGNs, spectroscopic
completeness drops dramatically at $z>1.2$, since at higher redshifts
the 4000\AA~break and the \OII~feature shift redward of the observed
wavelength range.  To ensure that X-ray objects with narrow-line and
lineless spectra are bona-fide AGNs, we select only objects with
$L_{0.5-10 {\rm keV}} > 3 \times 10^{42}$ erg~s$^{-1}$.  This X-ray
luminosity limit is generally used to separate AGNs from X-ray fainter
starburst galaxies \citep[e.g.,][]{hor01}.  We also include seven
broad-line AGNs without X-ray detection, six of which were selected by
their Spitzer/IRAC colors and one which is a serendipitous object from
the bright zCOSMOS survey (which selected targets based only on
$i_{\rm AB}<22.5$).  While these 7 X-ray undetected AGNs do not come
from a complete sample, we include them to gain a larger parameter
space of AGN spectral types and accretion rates (in effect, when using
their X-ray limits, they occupy the same $L_{disk}/L_X$ parameter
space as a few other X-ray detected AGNs).  Restricting narrow-line
and lineless AGNs to be X-ray luminous and adding the 7 X-ray
undetected broad-line AGNs makes a parent sample of 380 broad-line,
124 narrow-line, and 49 lineless AGNs (553 total) with high-confidence
redshifts and spectral identification.

Measuring accurate black hole masses additionally constrains the
sample to certain redshift ranges.  For Type 1 AGNs, we require the
presence of one of the \CIV, \MgII, or \Hb~broad emission lines in the
observed spectral range, effectively limiting broad-line AGNs with
IMACS or VIMOS spectra to $0.16<z<0.88$, $1<z<2.4$, and $2.7<z<4.9$
and objects with Hectospec or SDSS spectra to $z<4.9$.  For
narrow-line and lineless AGNs, we estimate black hole mass from the
$M_{BH} \sim L_{bulge}$ relation, and so we require an accurate
estimate of $L_{bulge}$.  For this we use the sample of objects in
COSMOS with morphological decompositions \citep{gab09} from the
HST/ACS images \citep{koe07}, which also effectively limits the
narrow-line and lineless AGNs to $z<1.2$ (beyond which the
4000\AA~break shifts out of the ACS-$i$ band and the host galaxy is
much more difficult to detect).  The accurate host measurements from
\citet{gab09} additionally allow us to subtract the host component
before computing the intrinsic bolometric luminosity.  In general, the
narrow-line and lineless AGNs are biased towards lower redshift and
consequently higher mass, since AGNs grow over the cosmic time.  The
narrow-line and lineless AGNs have a mean redshift of 0.7, while the
broad-line AGNs have a mean redshift of 1.6.  The final sample of 348
AGNs includes 256 broad-line, 65 narrow-line, and 27 lineless AGNs.

Full multiwavelength data exist for $>95\%$ of the AGNs in the sample
in every wavelength region except the UV.  X-ray data exist from both
Chandra and XMM-Newton: we use the deeper Chandra data when available,
but the Chandra observations cover only the central 0.8 deg$^2$ of the
COSMOS field.  For the 7 X-ray undetected broad-line AGNs, we use the
0.5-2~keV XMM flux limit ($f_{0.5-2 {\rm keV}} = 2 \times 10^{-16}$
erg~s$^{-1}$~cm$^{-2}$) for their X-ray luminosity (since these AGNs
have $L_{disk}/L_X>10$, their bolometric luminosity is dominated by
their optical/UV emission and completely neglecting their X-ray
emission does not significantly change their bolometric luminosity
estimate).  We apply the zero-point offsets derived by \citet{ilb09}
to the IR-UV photometry.

\begin{deluxetable*}{ccccrrrc}
\tablecolumns{7}
\tablewidth{0pc}
\tablecaption{COSMOS Multiwavelength Data
\label{tbl:obsdata}}
\tablehead{
  \colhead{Band} &
  \colhead{Telescope} & 
  \colhead{Wavelength} & 
  \colhead{Energy} &
  \colhead{Limit} &
  \colhead{NL/LL AGNs} &
  \colhead{BL AGNs} &
  \colhead{Reference\tablenotemark{a}} \\
  \colhead{} &
  \colhead{} &
  \colhead{\AA} &
  \colhead{eV} &
  \colhead{AB mag\tablenotemark{b}} &
  \colhead{Detected} &
  \colhead{Detected} &
  \colhead{} }
\startdata
X$_{hard}$ & Chandra & 1.24-6.20 & 2000-10000 & $7.3 \times 10^{-16}$ & 79/92 & 228/256 & (1) \\
X$_{hard}$ & XMM     & 1.24-6.20 & 2000-10000 & $9.3 \times 10^{-15}$ & 79/92 & 228/256 & (2) \\
X$_{soft}$ & Chandra & 6.20-24.8 & 500-2000   & $1.9 \times 10^{-16}$ & 88/92 & 249/256 & (1) \\
X$_{soft}$ & XMM     & 6.20-24.8 & 500-2000   & $1.7 \times 10^{-15}$ & 88/92 & 249/256 & (2) \\
FUV & GALEX & 1426-1667 & 7.44-8.63 & 25.7 & 27/92 & 131/256 & (3) \\
NUV & GALEX & 1912-2701 & 4.59-6.84 & 26.0 & 55/92 & 184/256 & (3) \\
u$^*$ & CFHT    & 3642-4180 & 2.97-3.40 & 26.4 & 92/92 & 254/256 & (4) \\
B$_J$ & Subaru  & 4036-4843 & 2.56-3.07 & 27.7 & 92/92 & 256/256 & (4) \\
g$^+$ & Subaru  & 4347-5310 & 2.33-2.85 & 27.1 & 92/92 & 256/256 & (4) \\
V$_J$ & Subaru  & 4982-5916 & 2.10-2.49 & 27.0 & 92/92 & 255/256 & (4) \\
r$^+$ & Subaru  & 5557-6906 & 1.80-2.23 & 27.1 & 92/92 & 256/256 & (4) \\
i$^*$ & CFHT    & 6140-9119 & 1.36-2.02 & 26.7 & 92/92 & 256/256 & (4) \\
F814W & HST/ACS & 7010-8880 & 1.40-1.77 & 27.2 & 92/92 & 256/256 & (5) \\
z$^+$ & Subaru  & 8544-9499 & 1.31-1.45 & 25.7 & 92/92 & 254/256 & (4) \\
J & UKIRT     & 11665-13223 & 0.94-1.06 & 23.8 & 92/92 & 256/256 & (4)\\
Ks & CFHT   & 19900-23050 & 0.538-0.623 & 23.4 & 92/92 & 253/256 & (6) \\
IRAC1 & Spitzer & 31557-38969 & 0.318-0.383 & 23.9 & 91/92 & 255/256 & (4) \\
IRAC2 & Spitzer & 39550-49663 & 0.250-0.313 & 23.3 & 91/92 & 255/256 & (4) \\
IRAC3 & Spitzer & 50015-63514 & 0.195-0.248 & 21.3 & 91/92 & 255/256 & (4) \\
IRAC4 & Spitzer & 62832-91229 & 0.136-0.197 & 21.0 & 91/92 & 255/256 & (4) \\
1.4 GHz & VLA & $2 \times 10^9$ & $6 \times 10^{-6}$ & 20$\mu$Jy & 92/92 & 256/256 & (7) \\
\enddata
\tablenotetext{a}{References are as follows: (1) \citet{elv09}, (2)
  \citet{cap09}, (3) \citet{zam07}, (4) \citet{cap10}, (5)
  \citet{koe07}, (6) \citet{mcc10}, (7) \citet{schi07}}
\tablenotetext{b}{X-ray flux limits are given in
  erg~s$^{-1}$~cm$^{-2}$, and the radio flux limit is given in
  $\mu$Jy.}
\end{deluxetable*}

%\subsection{Comparison Samples}
%
%Type 2 AGN with spectropolarimetry \citep{tran01,wang07}
%
%LINERs \& local AGNs \citep{ho09,era10}

\subsection{Measuring Absorption and Extinction}

X-ray absorption and optical/UV extinction could pose a challenge to
measuring the intrinsic accretion power.  The most heavily absorbed
AGNs (e.g. Compton-thick AGNs with $N_H>10^{24}$~cm$^{-2}$) are
entirely missed by our survey because they lack detectable X-ray
emission \citep[e.g.,][]{trei04}.  But if an AGN is moderately
absorbed and still X-ray detected, we might expect its disk to appear
cooler because the UV light is preferentially extincted, and its X-ray
slope to appear harder because the soft X-rays are preferentially
absorbed.  Some AGNs are also intrinsically reddened, decreasing their
UV emission by a factor of 2-3 \citep{ric03} and causing us to
underestimate their accretion disk emission.  With absorbed soft
X-rays and extincted disk emission, we could significantly
underestimate $L_{int}/L_{Edd}$.

We use X-ray column density $N_H$ to characterize the obscuration
properties of our AGNs.  Column density and optical extinction are
roughly correlated, with $A_V/N_H \sim 2 \times 10^{-23}$~cm$^2$
\citep{mar06}.  Then at $N_H<10^{22}$~cm$^{-2}$, optical magnitude
should be extincted by $\lesssim 20\%$ ($\lesssim 0.2$~mag).  Assuming
a SMC reddening law \citep{pei92}, as is most appropriate for AGNs,
this optical extinction translates to a factor of $\sim$1.2 extinction
at 3000\AA in the UV.  \citet{mai01} showed that the $A_V-N_H$
relation varies by up to a factor of 30 because of unknown changes in
the gas-to-dust ratio, grain size, and/or different physical locations
of the optical and X-ray absorbing material.  However for all AGNs in
the \citet{mai01} sample with $L_X>10^{42}$~erg~s$^{-1}$, $A_V/N_H <
1.8 \times 10^{-22}$~cm$^2$, meaning at $N_H \sim 10^{22}$~cm$^{-2}$
even the maximum optical ($V$-band) extinction is a factor of 5 and
the maximum UV (3000\AA) extinction a factor of 30.

\begin{figure}
\scalebox{1.2}
{\plotone{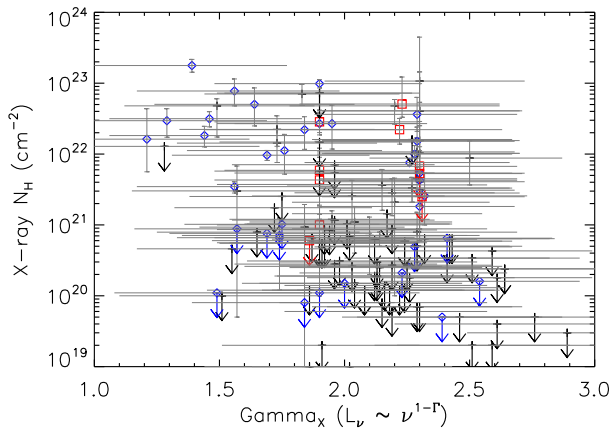}}
\figcaption{The column density $N_H$ and the X-ray slope $\Gamma_X$
  measured from the X-ray spectrum for the 153 AGNs with $>40$ XMM or
  Chandra counts.  X-ray slope $\Gamma_X$ is defined by $L_{\nu}
  \propto \nu^{1-\Gamma_X}$.  Black crosses show broad-line AGNs, blue
  diamonds show narrow-line AGNs, and red squares show lineless
  (optically dull) AGNs.  The median X-ray slope for all AGNs is
  $\Gamma_X=2.1$, although $\Gamma_X$ ranges from 1 to 3.  There are
  118 unobscured AGNs with $N_H<10^{22}$~cm$^{-2}$.
\label{fig:nhxslope}}
\end{figure}

Column density $N_H$ can be accurately measured for the 153 AGNs (93
broad-line, 38 narrow-line, 22 lineless AGNs) in the sample which have
$>40$ XMM or Chandra counts.  (With less than 40 counts, the spectral
fitting does not always stably converge.)  We fit each X-ray spectrum
as an intrinsically absorbed power-law with Galactic absorption
($N_{H,{\rm gal}} = 2.6 \times 10^{20}$~cm$^2$ in the direction of the
COSMOS field), with the power-law slope and $N_H$ as free parameters.
The best-fit $N_H$ value and its error are found using the
\citet{cash} statistic.  We present $N_H$ and X-ray slope $\Gamma$ in
Figure \ref{fig:nhxslope}.  Among the 153 AGNs with $>40$ X-ray
counts, there are 118 unobscured AGNs with $N_H<10^{22}$~cm$^{-2}$ (82
broad-line, 24 narrow-line, and 12 lineless AGNs).  We restrict our
main conclusions to this set of 118 unobscured AGNs for the remainder
of this work.

\section{Characterizing AGN Specific Accretion Rate}

In this work we describe the specific accretion rate using the
Eddington ratio parameter, $\lambda \equiv L_{int}/L_{Edd}$.  Here
$L_{int}$ is the intrinsic luminosity, a measure of the total
accretion luminosity which includes only light from the accretion disk
and X-ray corona and excludes any reprocessed IR emission.  While the
reprocessed IR emission can represent a large fraction of the
bolometric luminosity, especially for obscured AGNs, it may be
anisotropic.  Most of our AGNs are unobscured (see Section 2.1) and we
exclude the IR emission to avoid double-counting the AGN emission.
Instead we use only the optical/UV and X-ray emission which comes
directly from the disk and corona in the AGN: in this work when using
``intrinsic'' luminosity we are always referring to the total of the
disk (optical/UV) and corona (X-ray) emission, without the reprocessed
(IR) emission.  The Eddington luminosity is derived from the black
hole mass, with $L_{Edd}=1.3 \times 10^{38}
(M_{BH}/M_{\odot})$~erg~s$^{-1}$.  AGN luminosity is powered by
accretion rate, with $L_{int} = \eta \dot{M} c^2$.  For a constant
efficiency $\eta$, the Eddington ratio $\lambda$ is equivalent to the
specific accretion rate $\dot{m} \equiv \dot{M}/\dot{M_{Edd}}$.  For
example, assuming $\eta \sim 0.1$ the Eddington accretion rate can be
written $\dot{M}_{Edd} = 5M_8$~M$_{\odot}$~yr$^{-1}$ with
$M_8=M/(10^8M_{\odot})$.  However there is good evidence that $\eta$
decreases at very low accretion rates $\dot{m} << 0.01$
\citep[e.g.,][]{nar08}.  Indeed, in Sections 4 and 5 we invoke a
lower-efficiency (radiatively inefficient) accretion to explain the
observational properties of $L_{int}/L_{Edd} < 10^{-2}$ AGNs.  This
means that the accretion power $L_{int}/L_{Edd}$ probably
underestimates the accretion rate $\dot{m}$ for our most weakly
accreting AGNs with $L_{int}/L_{Edd} < 10^{-2}$: e.g., a measured
accretion power of $L_{int}/L_{Edd} \sim 10^{-4}$ might correspond to
$\dot{m} \sim 10^{-3}$.

Below we outline our methods for estimating black hole masses and
bolometric luminosities from the data for the AGNs in our sample.
Table \ref{tbl:catalog} presents the full catalog of $L_{int}$,
$M_{BH}$, and $L_{int}/L_{Edd}$, and their associated errors, for our
AGNs.

\subsection{Intrinsic Luminosity Estimates}

We calculate the intrinsic luminosity from the full rest-frame near-IR
to X-ray multiwavelength data.  This avoids monochromatic bolometric
corrections which are highly uncertain and probably depend on
Eddington ratio \citep[e.g.,][]{kel08, vas09}.  Instead we measure
intrinsic luminosity by integrating the best-fit accretion disk +
X-ray power-law SED model.  We compile the broad-band near-IR ($Ks$,
$J$), optical ($z^+$, $r^+$, $i^*$, $g^+$, $V_J$, $B_J$, $u^*$), UV
(GALEX NUV \& FUV), and X-ray (0.5-2 keV and 2-10 keV from Chandra
when available or XMM-Newton) data, for which the wavebands and limits
are described in Table \ref{tbl:obsdata}.  To avoid reprocessed mid-IR
emission, which would double-count the intrinsic emission for an
unobscured AGN, we restrict the accretion disk fit to rest-frame
$1<E<100$~eV ($6200>\lambda>124$~\AA).  The radio emission is
negligible in the total energy output of our AGNs.  While narrow-band
optical photometry also exists for our AGNs, its inclusion doesn't
appreciably change the best-fit multiwavelength SED from using only
the broad-band data.

The rest-frame near-IR and optical emission of narrow-line and
lineless AGNs is dominated by the emission from the host galaxy.  For
these objects, accurate intrinsic luminosities require modeling and
subtracting the host galaxy light.  \citet{gab09} measured the host
F814W luminosities from surface brightness fitting to the HST/ACS data
of our AGNs.  We use this luminosity to scale a galaxy template from
\citet{pol07}.  Lineless AGNs have early-type hosts, since their
spectra lack the emission lines associated with a late-type
star-forming galaxy, and so we use the ``Ell5'' early-type template
from \citet{pol07}.  The narrow-line galaxies in our sample typically
have intermediate-type (``green valley'') hosts based on their
morphologies \citep{gab09} and star formation rates \citep{sil09}, and
so we use the ``S0'' template of \citet{pol07}.  We subtract the host
contribution in each photometric band before performing our SED fit.
The reddest (``Ell2'') and bluest (``Sd'') normal galaxy templates of
\citet{pol07} are additionally used as extreme hosts to estimate the
possible error contribution from choosing the wrong host template
(described in \S 3.3).

It is possible that a few of the narrow-line and lineless AGNs might
have very blue starbursting hosts, although such galaxies are uncommon
at $z<1$.  An extremely blue, UV-emitting host would cause us to
overestimate the accretion disk emission and consequently overestimate
the accretion rate.  Since the narrow-line and lineless AGNs have
lower accretion rates than broad-line AGNs (as we discuss in Section
4), if their true accretion rates were even lower it would only
strengthen our conclusions.  It is also possible that very red, dusty
hosts could cause us to underestimate the true accretion rates for
narrow-line and lineless AGNs.  However a dusty host should cause the
AGN to appear extincted, and our sample of AGNs generally has low
measured absorption (see Section 2.1).  In addition, restricting our
fitting to $1<E<100$~eV ($6200>\lambda>124$~\AA) already means that a
normal elliptical galaxy (like our ``Ell2'' template) contributes very
little flux where we fit the accretion disk.

While broad-line AGNs are likely to have some host contribution, we
cannot use surface brightness fitting to estimate their host
luminosity because they are at high redshift and their point source
overwhelms their extended emission \citep{gab09}.  However at the peak
of the accretion disk emission for a broad-line AGN ($\sim3000$\AA, or
4~eV) the host galaxy contributes $<20$\% of the emission
\citep[e.g.,][]{ben06}.  Because we additionally restrict our
accretion disk fitting to $1<E<100$~eV ($6200>\lambda>124$~\AA), we
can assume that the error from not subtracting the host for broad-line
AGNs is typically $<0.1$dex.

\begin{figure*}
\centering
\scalebox{1.1}
{\plotone{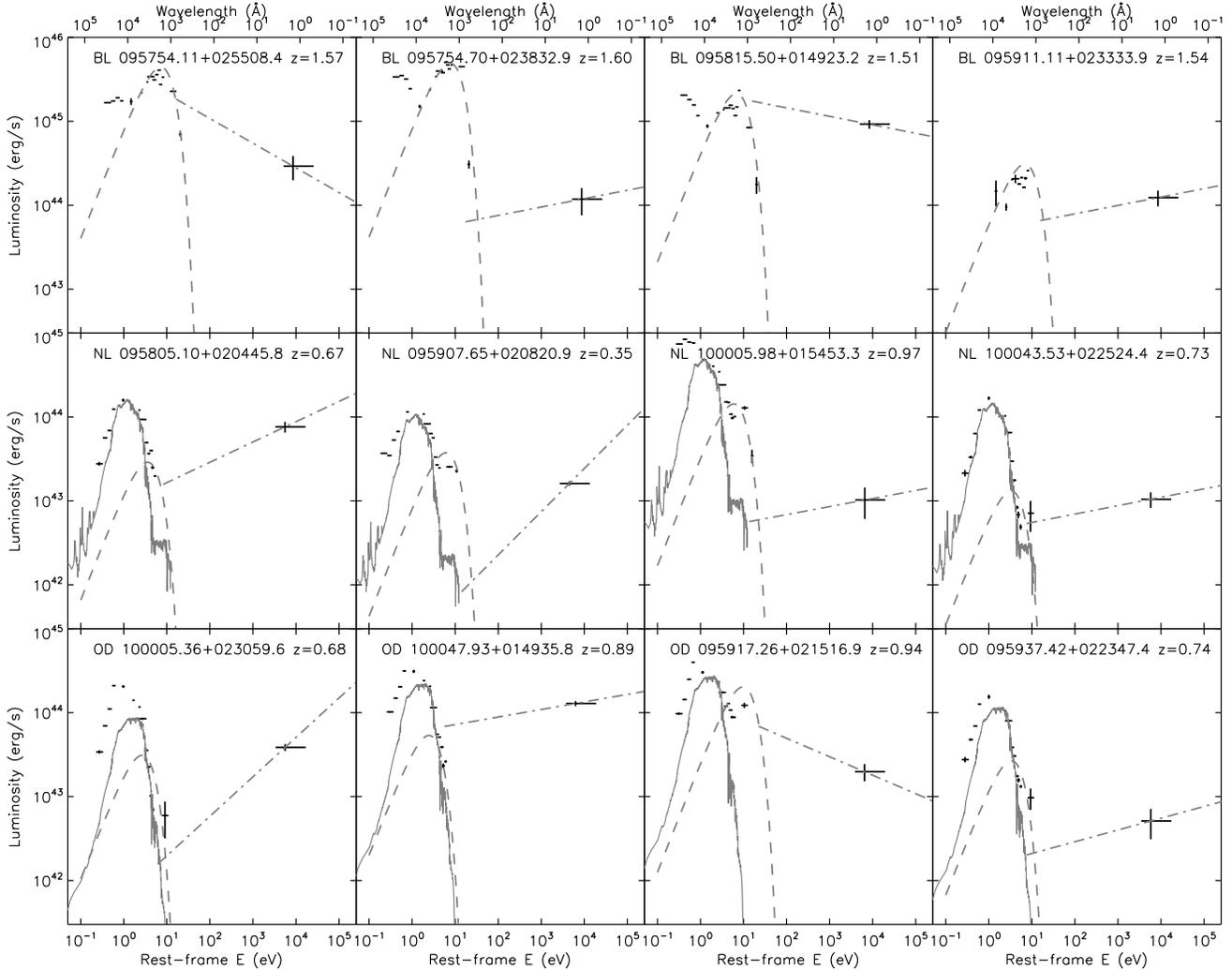}}
\figcaption{Multiwavelength photometry and model fits for 12 example
  AGNs.  The top four panels are broad-line AGNs (represented by
  'BL'), the middle four are narrow-line AGNs (represented by 'NL'),
  and the bottom four are lineless ``optically dull'' AGNs
  (represented by 'OD').  In each panel, the dashed line is the
  best-fit accretion disk model and the dot-dashed line is the X-ray
  power-law fit.  The X-ray power-law slope comes from the X-ray
  spectral fit, although we show only the X-ray photometry data in
  this figure.  Estimated host SEDs are shown by solid lines for the
  narrow-line and lineless AGNs.  We fit only at $E>1$~keV in order to
  ignore the reprocessed IR emission, and so the longest-wavelength
  photometry data (especially the IRAC channels) are not fit by our
  models.
\label{fig:sedfits}}
\end{figure*}

We shift the observed (and host-subtracted, for narrow-line and
lineless AGNs) photometry to the rest-frame from the measured
spectroscopic redshift and convert the magnitudes or fluxes to
luminosities.  We then fit an accretion disk model to the optical/UV
emission within the range $1<E<100$~eV ($4.8 \times 10^{14}<\nu<2.4
\times 10^{16}$~Hz, or $6200>\lambda>124$~\AA) and a power-law
representing the X-ray corona emission to the rest-frame X-ray data.
We measure the total bolometric luminosity from the sum of the disk
luminosity (given by the analytic solution in Equation 3 below) and
the power-law luminosity from $4E_{peak}<E<250$~keV (where $E_{peak}$
is the peak energy of the best-fit disk model).  While the X-ray
background requires a high-energy cutoff for AGNs in the few hundreds
of keV \citep{gil07}, measurements of the cutoff energy exist for only
$\sim$15 AGNs and vary from 50-500 keV \citep{per02,mol06}.  We choose
250 keV as an intermediate value, although any cutoff from 50-500 keV
does not greatly influence our results.  Our AGNs have typically flat
X-ray spectra with $\Gamma_X \sim 2$, and so changing the X-ray cutoff
energy by a factor of 0.2-2 effectively changes the integrated X-ray
luminosity by the same factor of a few.  Because the X-ray and disk
luminosities are roughly comparable (see Figure \ref{fig:sedfits}),
this results in less than a factor of two change in the total
accretion luminosity: much less than the $\sim$0.5~dex errors we
compute for our estimated $L_{int}/L_{Edd}$ (see \S 3.3).

We use the accretion disk model of \citet{gier99}, which improves upon
a basic blackbody accretion disk by including a correction for
relativistic effects.  (The \citet{gier99} model is the ``diskpn''
model of the XSpec X-ray fitting software.)  This model is based on
the pseudo-Newtonian gravitational potential $\Phi=-GM/(R-R_g)$
\citep{pac80}, where $R_g$ is the Schwarzschild radius $R_g=GM/c^2$.
From \citet{gier99}, the model takes the form:
\begin{equation}
  L = KE^4 \int_{r_{in}}^{\infty}{\frac{r{\rm d}r} {exp[E/kT(r)]-1}}
\end{equation}
where $r=R/R_g$ and we assume the innermost stable orbit $r_{in}=6$.
The temperature depends on radius as
\begin{equation}
  T(r) = \frac{T_0}{c_0} \left[ \frac {r-2/3} {r(r-2)^3}
           \left(1 - \frac{3^{3/2}(r-2)}{2^{1/2}r^{3/2}} \right) \right]^{1/4},
\end{equation}
with $c_0 \simeq 0.1067$, and $T_0 \propto m^{-1/4}\dot{m}^{1/4}$.  The
coefficient $K$ depends on inclination angle, coronal absorption, and
the color to effective temperature ratio.  Rather than estimate these
values, we assume that $K$ is a constant, computed by simply scaling
the model to our data.  $T_0$ is the sole free parameter.  In our
analyses below we refer to $E_{peak}$, the peak energy of the disk,
rather than $T_0$, and in general $kT_0 \simeq E_{peak}/24$.  We find
the best-fit disk model in terms of $T_0$ by minimizing the $\chi^2$
function.  While most of the best-fit disk models have significant
emission at $E<1$~eV, we restrict the fit to $1<E<100$~eV to mitigate
the effects of a contaminating torus and/or host galaxy light.

Note that the relation $T_0 \propto m^{-1/4}\dot{m}^{1/4}$ above means
that the disk temperature is constrained not only by the photometry by
also by the black hole mass.  In practice this prevents our fits from
resulting in unphysically hot accretion disks, since disks peaking at
energies much higher than $\sim$4~eV (3000\AA) would require
unphysically small black hole masses.  This is especially important to
note because about one-third of the sample lacks GALEX UV detections,
and as a result the declining high-energy slope is not well
constrained by the photometry for low redshift AGNs.  The black hole
mass error ($\sim0.4$~dex) is used during the bootstrapped uncertainty
measurements for the accretion disk temperature and luminosity.

The total disk luminosity is calculated analytically \citep[see
  Appendix A of][]{gier99}:
\begin{equation}
  L_{disk} = K \frac{h^3c^2}{16\pi} \left(\frac{T(r_{in})}{c_0}\right)^4.
\end{equation}
Errors in both $E_{peak}$ and $L_{disk}$ are found by bootstrapping
1000 fits to the resampled data.

To characterize the X-ray corona emission, we use the X-ray spectral
fits described in \S 2.1.  Each X-ray spectrum is fit as an
intrinsically absorbed power-law with Galactic absorption ($N_{H,{\rm
gal}} = 2.6 \times 10^{20}$~cm$^2$ in the direction of the COSMOS
field).  We use the photon index $\Gamma_X$ to represent the power-law
slope, such that $L_{\nu}=L_0\nu^{1-\Gamma}$.  Figure
\ref{fig:nhxslope} shows that the typical $\Gamma_X \simeq 1.9 \pm
0.4$, and we assume this slope for AGNs with too few X-ray counts for
a good fit.  We calculate the total X-ray luminosity by integrating
the power-law model over $4E_{peak}<E<250$~keV (where $E_{peak}$ is
the energy peak of the disk model), using the analytic solution:

\begin{equation}
  L_X = L_0/(2-\Gamma) \times [(250{\rm keV}/h)^{2-\Gamma}-(4E_{peak}/h)^{2-\Gamma}]
\end{equation}

The total bolometric luminosity is simply the sum of the integrated
accretion disk and X-ray power-law components, $L_{int}=L_{disk}+L_X$.

Figure \ref{fig:sedfits} shows a representative sample of broad-line,
narrow-line, and lineless SEDs with model fits.  Note that emission
lines and variability (the various photometric data were taken over 3
years) mean that our simple accretion disk model is not a perfect fit:
some of the optical/UV data and differ from the model fit by up to
0.2~ dex.  However, such small errors in individual photometry points
are negligible compared to the $>0.4$~dex total errors we estimate for
$L_{int}$ (see Figure \ref{fig:lbolmbh} and Section 3.3).  In general,
the accretion disk plus X-ray power-law model provides an accurate,
physically motivated fit to the data.

\subsection{Black Hole Mass Estimates}

For Type 1 AGNs, we estimate black hole masses using the scaling
relations of \citet{ves09} for the \MgII~broad emission line and
\citet{ves06} for the \Hb~and \CIV~broad emission lines.  These
relations estimate black hole mass from single-epoch spectra by
employing the correlation between the radius of the broad emission
line region and the continuum luminosity, $R_{BLR} \sim L^{0.5}$,
observed in local AGN with reverberation mapping \citep{ben06, kas07}.
In general, masses estimated from the scaling relations are accurate
to $\sim 0.4$ dex \citep{ves06, shen08} and agree with local AGN
masses from dynamical estimators \citep{dav06, onk07} and the
$M_{BH}$-$\sigma*$ correlation \citep{onk04, gre06}.  The scaling
relations take the form of Equation 5, with ${\lambda}L_{\lambda}$ in
units of $10^{44}$~erg/s and $v_{FWHM}$ in units of 1000~km/s;
$A=6.91$, $B=0.50$, and $\lambda=5100$\AA~for \Hb; $A=6.86$, $B=0.50$,
and $\lambda=3000$\AA~for \MgII; $A=6.66$, $B=0.53$, and
$\lambda=1350$\AA~for \CIV.
\begin{equation}
  \log \left( \frac{M_{BH}}{M_{\odot}} \right) = 
  A + B \log(\lambda L_{\lambda}) + 2 \log(v_{FWHM})
\end{equation}

Black hole masses for the Type 1 AGNs with Magellan/IMACS or SDSS
spectra in COSMOS are already published in previous work
\citep{tru09b}, and we repeat the same techniques for Type 1 AGNs with
VLT/VIMOS spectra.  Briefly, a power-law fit plus iron emission are
fit to each AGN.  The continuum luminosity is estimated directly from
the continuum fit, while the velocity widths are computed from
Gaussian fits to the continuum-subtracted emission lines.  Some
objects also have black hole masses from \citet{mer10}; for these
objects, our masses are consistent with a random scatter of only
$\sim$0.4 dex: equivalent to the intrinsic scatter of the scaling
relations \citep[see Figure 3 of][]{tru09b}.  \citet{mar08} showed
that the scatter in $M_{BH}$ from the scaling relations might decrease
to 0.2 dex if radiation pressure is taken into account.  Replacing the
scaling relations from Equation 5 with those of \citet{mar08} would
tighten the distribution of $L_{int}/L_{Edd}$ estimates for broad-line
AGNs about $L_{int}/L_{Edd} \sim 0.3$.  This has no impact on the
$M_{BH}$ estimates for narrow-line and lineless AGNs, and does not
affect the difference in $L_{int}/L_{Edd}$ between the broad-line
sample and the narrow-line and lineless AGN sample.

Estimating black hole masses for AGNs without broad emission lines
requires secondary estimators.  We employ the relationship between
$M_{BH}$ and rest-frame $K$-band host bulge luminosity \citep{gra07}:
\begin{equation}
  \log \left( \frac{M_{BH}}{M_{\odot}} \right) = 0.93 (\log(L_K)-0.3z) - 32.30,
\end{equation}
with $L_K$ in units of erg~s$^{-1}$.  The $M_{BH}-L_{K,bulge}$
relation comes from the more fundamental $M_{BH}-M_*$ relation, since
rest-frame $K$ bulge luminosity is correlated with $M_*$
\citep[e.g.,][]{ilb10}.  We add an additional $-0.3z$ term to the
\citep{gra07} relation in order to account for the evolution in the
$M_*/L_K$ ratio, $\log(M_*/L_K) \propto -0.3z$ \citep{arn07}.  We
measure rest-frame $L_K$ from the host galaxy template from the
multiwavelength SED fit (described above in \S 3.1).  The early-type
template for the lineless AGNs is, by definition, bulge-dominated, and
so $L_{K,bulge} = L_{K,host}$.  The S0 template used for the
narrow-line AGNs, however, has a significant disk component, and so we
take $L_{K,bulge} = 0.5L_{K,host}$.  The intrinsic error in the
$M_{BH}-L_K$ is 0.35 dex \citep{gra07}.  We do not correct the
$M_{BH}$ estimates for any evolution in the $M_{\rm bulge}-M_{BH}$
relation because measuring of $M_{\rm bulge}-M_{BH}$ evolution has
proved difficult due to significant biases in most tests
\citep{lau07,shen10}.  Besides, while there is some evidence for
evolution to $z \sim 3$ \citep{dec10}, there is probably little or no
evolution to $z \sim 1.5$ \citep{jah09} and our narrow-line and
lineless AGNs lie at $z<1$.

\begin{figure}
\scalebox{1.2}
{\plotone{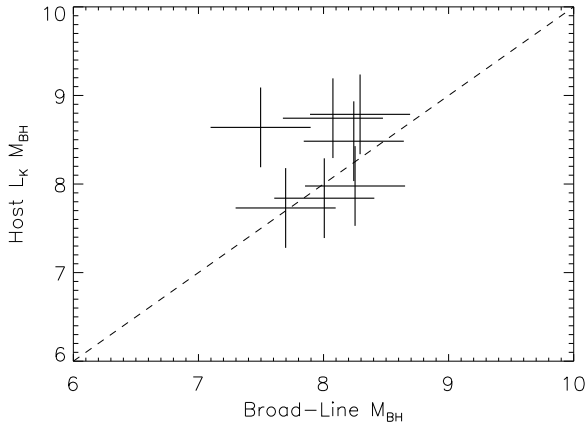}}
\figcaption{Black hole mass estimates from both the host $L_K$ and the
  broad-line scaling relations for the seven broad-line AGNs with
  detected host galaxies from \citet{gab09}.  For all but one AGN,
  both $M_{BH}$ estimates agree within $\lesssim 1\sigma$ (for the
  remaining object, the two estimates differ by only $\sim 2\sigma$).
  From these AGNs, and the sets of nearby AGNs with similarly
  consistent masses from both estimators \citep{onk04, gre06}, it is
  unlikely that the different mass estimators cause bias between
  broad-line and narrow-line/lineless AGNs.
\label{fig:checkmbh}}
\end{figure}

Because we use different mass estimators for broad-line and
narrow-line/lineless AGNs, it is important to demonstrate that the two
methods agree.  Seven of our broad-line AGNs have detected host
galaxies from the decompositions of \citet{gab09}, and for these AGNs
we compare $M_{BH}$ estimates from the broad-line scaling relations
and from the host galaxy rest-frame $L_K$ in Figure
\ref{fig:checkmbh}.  The $M_{BH}$ estimates from broad lines and $L_K$
agree within $<2\sigma$ for all objects (indeed, estimates for all
objects but one agree within $<1\sigma$).  In addition to the seven
broad-line AGNs in our sample, both the broad-line and host galaxy
$M_{BH}$ estimators have been shown to produce consistent masses for
nearby AGNs \citep{onk04, gre06}.  It is particularly unlikely that
either of the estimators is systematically off by a factor of 100.
Therefore we are confident that the factor of 100 difference in
$L_{int}/L_{Edd}$ for broad-line and narrow-line/lineless AGNs in
Section 4 (see, for example, Figure \ref{fig:acchistogram}) is a
physical effect, robust beyond the choice of black hole mass
estimator.

\begin{figure}
\scalebox{1.2}
{\plotone{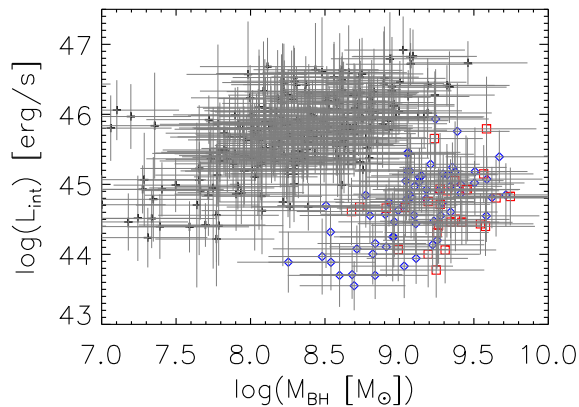}}
\figcaption{Intrinsic luminosity $L_{int}$ with black hole mass
  $M_{BH}$ for the AGN sample.  Broad-line AGNs are shown by black
  crosses, narrow-line AGNs by blue diamonds, and lineless AGNs by red
  squares.  Errors are calculated as described in Section 3.3.
  Narrow-line and lineless AGNs generally have higher mass, due to the
  COSMOS selection limits, but they also have lower luminosities as
  expected by downsizing.
\label{fig:lbolmbh}}
\end{figure}

We highlight the range and limitations of the AGN sample in Figure
\ref{fig:lbolmbh}, which shows bolometric luminosities and black hole
masses for the broad-line, narrow-line, and lineless AGNs.  Objects in
the upper left have the highest specific accretion rates, while those
in the lower right are weakly accreting AGNs.  While the total sample
spans 3 orders of magnitude in both luminosity and black hole mass,
our narrow-line and lineless AGNs are generally less luminous and more
massive than broad-line AGNs.  The lack of low-mass narrow-line and
lineless AGNs is due to the selection limits of the survey: such
objects are too faint to be detected in COSMOS.  It is suggestive that
these higher mass narrow-line and lineless AGNs are at $z<1$ and are
less luminous: this is consistent with '' downsizing,'' with more
massive AGNs becoming less active at lower redshift
\citep{ued03,bra05,bon07}.

Figure \ref{fig:lbolmbh} shows that at a given mass or luminosity
there are generally all types of AGNs present in our sample.  For this
reason we do not expect that the differences between broad-line and
narrow-line/lineless AGNs are biased by selected samples from
different masses or luminosities.  In addition, despite the different
redshifts of most broad-line and narrow-line/lineless AGNs, we do not
expect their differences to be caused by redshift.  There is evidence
that AGN obscuration properties depend on redshift \citep{trei09,
  tru09a}, but these AGNs are unobscured.  The AGN central engine,
meanwhile, does not change with redshift in terms of ionization
parameters \citep{die04,ves04}, spectral energy distributions
\citep{vig03,ric06,kel08}, or metallicity \citep{sim10}.  Limiting the
sample to $z<1$, $8.5<\log(M_{BH})<9$, or $44<\log(L_{int})<45$ does
not significantly change the differences between the broad-line and
narrow-line/lineless AGN samples seen in Figures 5, 6, 7, or 8.

\subsection{Error Budget}

We estimate errors for each of our specific accretion rates,
propagating the errors from both the intrinsic luminosity estimate and
the black hole mass estimate.  Our intrinsic luminosity is subject to
three major uncertainties:
\begin{itemize}
\item Photometry errors, $\sigma_{phot}$.  We measure the error
  contribution of the photometry by bootstrapping, fitting our model
  SED to 1000 realizations of randomly drawn photometry values
  distributed according to the measurement errors.  In general,
  $\sigma_{phot} \sim 0.1$~dex.
\item Errors in the host subtraction, $\sigma_{host}$.  For broad-line
  AGN, we do not subtract a host component and assume that any
  remaining galaxy light overestimates the intrinsic luminosity (from
  the UV and X-ray) by only $<0.1$dex (see \S 3.1).  For narrow-line
  and lineless AGN we estimate $\sigma_{host}$ from the difference in
  the resultant $L_{int}$ when using a very red (``Ell2'') and a very
  blue (``Sd'') template from \citet{pol07}.  Since the accretion disk
  is fit only at $E>1$~eV where there is little host emission (even
  from the ``Sd'' galaxy), this error is usually insignificant
  ($\sigma_{host} \lesssim 0.1$~dex).
\item Incorrect $L_{int}$ resulting from extinction, $\sigma_{ext}$.
  Extinction will make the true $L_{disk}$ greater than our estimate
  because optical/UV light will be missed, but will make the true
  $L_X$ lower than our estimate because the power-law slope will be
  too hard.  Because we restrict our main analyses to unobscured
  ($N_H<10^{22}$~cm$^{-2}$) AGNs, we assume this error is $<0.1$~dex
  (see \S 2.1).
\end{itemize}

The black hole estimate is subject to two major uncertainties:
\begin{itemize}
\item Intrinsic errors in the $M_{BH}$ relations, $\sigma_{rel}$.  For
  broad-line AGN, the intrinsic error in the scaling relations is
  0.4~dex \citep{ves06}, such that $\sigma_{rel}=2.5M_{BH}$.  For
  narrow-line and lineless AGN, we use the $M_{BH}-L_{K,host}$
  relation, and its associated intrinsic scatter is $0.35$~dex
  \citep{gra07}, such that $\sigma_{rel}/M_{BH}=2.2$.  These errors
  dominate the error in $L_{int}/L_{Edd}$, except for highly absorbed
  AGNs with $N_H>10^{22.5}$~cm$^{-2}$.
\item Measurement error in the luminosity used in the scaling
  relation, $\sigma_{lum}$.  For broad-line AGN, this is the measured
  continuum luminosity associated with the appropriate scaling
  relation, estimated by \citet{tru09b} as $\sigma_{lum} \sim
  0.05$~dex.  Since $M_{BH} \propto L^{0.5}$, $\sigma_{lum}=1.3M_{BH}$
  for broad-line AGNs.  For other AGNs the $\sigma_{lum}$ comes from
  our measured $L_{K,rest}$.  We estimate this error for the
  narrow-line and lineless AGNs from 1000 fits to the randomly
  subsampled data, and find that the error is generally insignificant
  compared to the intrinsic error ($\sigma_{lum} \sim 0.05$~dex).
  Note the contribution from error in $v_{FWHM}$ to $M_{BH}$ in
  broad-line AGNs is also negligible, since for our AGNs
  $\sigma(v_{FWHM}) < 0.2 v_{FWHM}$ \citep{tru09b}.
\end{itemize}

The total error in specific accretion rate, $\sigma_{\lambda}$, is
then given by:
\begin{equation}
  \frac{\sigma_{\lambda}^2}{\lambda^2} = 
    \frac{ \sigma_{phot}^2 + \sigma_{host}^2 } {L_{int}^2} 
    + \frac{\sigma_{rel}^2 + \sigma_{lum}^2} {M_{BH}^2}
\end{equation}
We measure the total error $\sigma_{\lambda}$ by bootstrapping, with
1000 fits to the resampled data.  In each fit we allow all of the
parameters above to vary according to their error.  The intrinsic
error in the $M_{BH}$ relations ($\sigma_{rel}$) dominates the error.
The average errors are $\sim$0.5 dex, compared to the $\sim$4 dex
range in $L_{int}/L_{Edd}$ for the AGN in the sample.

\section{The Physical Effects of Specific Accretion Rate}

\begin{figure}
\scalebox{1.2}
{\plotone{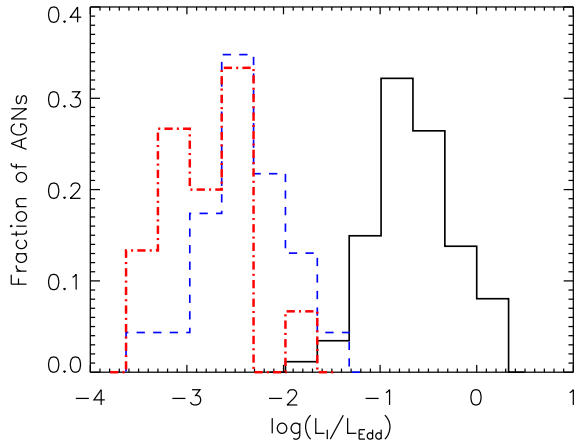}}
\figcaption{The distribution of calculated specific accretion rates
  ($L_{int}/L_{Edd}$), for the 82 unobscured ($N_H<10^{22}$~cm$^{-2}$)
  broad-line AGNs (black histogram), 24 narrow-line AGNs (blue dashed
  histogram), and 12 lineless AGNs (red dotted histogram).
  Narrow-line and lineless AGNs have significantly lower accretion
  rates than broad-line AGNs.  The $L_{int}/L_{Edd} \gtrsim 0.01$
  limit for broad-line AGNs is not a selection effect \citep{tru09b}.
\label{fig:acchistogram}}
\end{figure}

The distribution of $L_{int}/L_{Edd}$ for the 118 unobscured AGNs is
shown in Figure \ref{fig:acchistogram}.  It is immediately evident
that unobscured narrow-line and lineless AGNs accrete much more weakly
than broad-line AGNs, with specific accretion rates differing, on
average, by $\sim$2 orders of magnitude.  This suggests that many
narrow-line and lineless AGNs are not simply geometrically obscured
versions of broad-line AGNs, but they instead have fundamentally
different accretion physics which we examine in more detail below.

The large $\sim$0.5 dex errors in accretion rate artificially broaden
the distributions, such that the intrinsic distributions are likely to
be narrower than the histograms in Figure \ref{fig:acchistogram}
appear (although many $L_{int}/L_{Edd} \lesssim 10^{-3}$ narrow-line
and lineless AGNs could be too faint for the COSMOS X-ray and
spectroscopy limits).  The $L_{int}/L_{Edd} \gtrsim 0.01$ limit for
broad-line AGNs could be partially explained by selection effects
\citep{kel10}, since low accretion rates AGNs are typically less
luminous.  However at the highest masses ($M_{BH} \sim 10^9
M_{\odot}$), broad-line AGNs with $L_{int}/L_{Edd} \lesssim 0.01$ must
be very rare \citep{kol06,tru09b}.  Meanwhile unobscured narrow-line
and lineless AGNs are generally limited by $L_{int}/L_{Edd} \lesssim
0.01$.  With low X-ray column densities and low accretion rates, these
objects have similar properties to the ``naked'' Type 2 AGNs of
\citet{tran03}, which additionally lack reflected broad emission lines
in spectropolarimetry \citep[see also][]{gli07, wang07}.  We expect
that the X-ray unobscured low accretion rate AGNs would similarly lack
reflected broad emission lines.  Our method cannot accurately estimate
$L_{int}/L_{Edd}$ for obscured AGNs, but following a unified model
with geometric obscuration \citep[e.g.,][]{ant93}, obscured narrow-line
AGNs would likely have accretion rates comparable to our broad-line
AGNs.

%% For comparison we also estimate $L_{int}/L_{Edd}$ for 14 obscured
%% narrow-line AGNs with $>40$ X-ray counts and $N_H>10^{22}$~cm$^{-2}$.
%% We cannot use the method of \S 3.1 to estimate the intrinsic accretion
%% luminosity of these obscured AGNs.  Instead we use a bolometric
%% correction from the reprocessed $6 \mu$m emission, $L_{int} \simeq
%% 8L_{6 \mu {\rm m}}$ \citep{ric06}.  We then estimate $M_{BH}$ from the
%% host luminosity in the same manner described in \S 3.2 and show the
%% distribution of $L_{int}/L_{Edd}$ for these obscured AGNs as the
%% filled histogram in Figure \ref{fig:acchistogram}.  The measure of
%% $L_{int}/L_{Edd}$ for the obscured AGNs is highly uncertain:
%% \citet{ric06} formally measure the bolometric correction as $8 \pm 3$
%% in their sample of luminous quasars, and application to these obscured
%% narrow-line AGNs probably leads to even larger uncertainties.  However
%% it is instructive to note that the obscured narrow-line AGNs have
%% roughly the same accretion rates as the broad-line AGN population,
%% while unobscured narrow-line AGNs lower $L_{int}/L_{Edd}$.

\begin{figure*}
\centering
\scalebox{1.1}
{\plotone{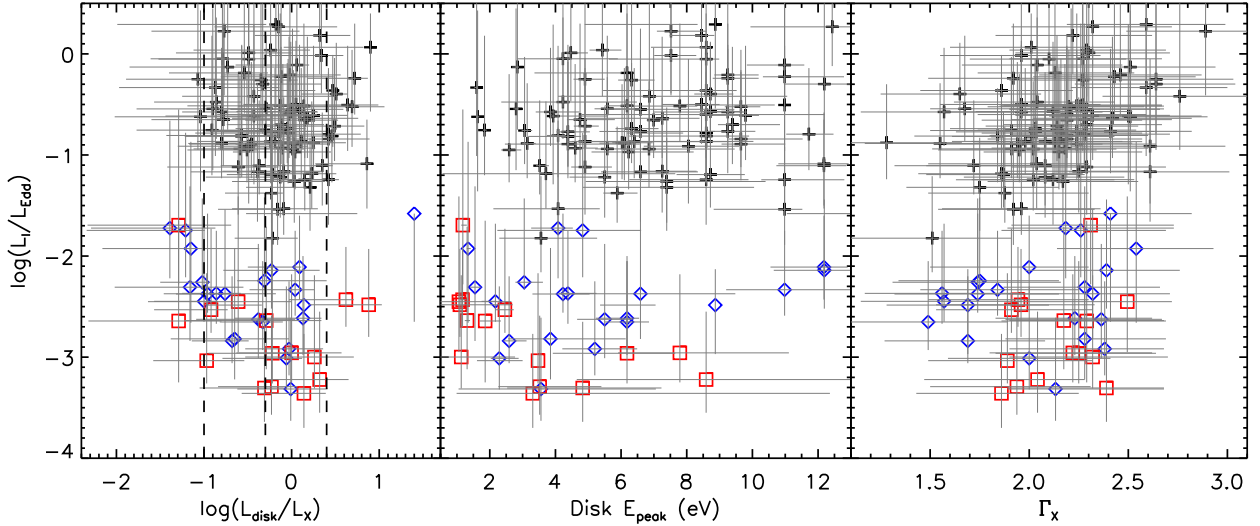}}
\figcaption{Specific accretion rate $L_{int}/L_{Edd}$ and the ratio of
  disk to corona emission $\log(L_{disk}/L_{X})$, disk temperature
  $E_{peak}$ and X-ray photon index $\Gamma_X$ for the 118 unobscured
  AGNs with $N_H<10^{22}$~cm$^{-2}$.  In each panel, black crosses
  represent broad-line AGNs, blue diamonds are narrow-line AGNs, and
  red squares are lineless AGNs.  The dashed lines in the left panel
  show lines of $\alpha_{OX}=1.0, 1.5, 2.0$, assuming $E_{peak}=6$~keV
  and $\Gamma_X=1.9$.  Unobscured narrow-line and lineless AGNS have
  $\sim$100 times lower accretion rates than broad-line AGNs, as well
  as significantly cooler and somewhat weaker accretion disks.
\label{fig:accretion}}
\end{figure*}

We can compare the specific accretion rates and AGN types with the
physical parameters of our model fits: namely the ratio of disk to
power-law emission, the peak energy of the accretion disk model, and
the X-ray power-law slope.  These quantities are particularly useful
in unifying AGN in terms of their accretion physics.  Figure
\ref{fig:accretion} shows the specific accretion rate with these
parameters for each AGN type.  The values of $L_{disk}/L_X$ can be
roughly translated to values of $\alpha_{OX}$, with
$\alpha_{OX}=-0.384\log[L_{\nu}(2500\AA)/L_{\nu}(2{\rm keV})]$
\citep{tan79,kel08}.  The left panel of Figure \ref{fig:accretion}
shows tracks of $\alpha_{OX} = 1.0, 1.5, 2.0$, assuming
$E_{peak}=6$~keV and $\Gamma_X=1.9$ (hotter disks and softer X-ray
slopes increase $\alpha_{OX}$).  Once again, narrow-line and lineless
AGNs have lower specific accretion rates, and they also tend to have
lower $L_{disk}/L_X$ and $E_{peak}$.

We can determine the significance of any differences in
$L_{disk}/L_X$, $E_{peak}$, and $\Gamma_X$ between rapidly accreting
broad-line AGNs and weakly accreting narrow-line and lineless AGNs by
comparing their mean values and considering the scatter of each
sample.  Given mean values $\mu_1$ and $\mu_2$ and associated scatters
$\sigma_1$ and $\sigma_2$ for each set, the significance of their
difference is given by $(\mu_1-\mu_2) / \sqrt(\sigma_1^2/N_1 +
\sigma_2^2/N_2)$ (where $N_1$ and $N_2$ are the numbers of AGNs in
each sample.  The broad-line AGNs have $\mu(\log(L_{disk}/L_X)) =
-0.14 \pm 0.44$ while the narrow-line and lineless AGNs have
$\mu(\log(L_{disk}/L_X)) = -0.38 \pm 0.64$, so that their difference
is marginally significant at $2.1\sigma$.  The difference in
$E_{peak}$ is more significant: the broad-line AGNs have
$\mu(\log(E_{peak})) = 0.80 \pm 0.20$ and the narrow-line/lineless
AGNs have $\mu(\log(E_{peak})) = 0.59 \pm 0.37$, so that the
difference is significant to $3.3\sigma$.  From this we can conclude
that a transition from weakly accreting narrow-line and lineless AGNs
to the rapidly accreting broad-line AGNs results in significantly
hotter and marginally brighter emission from the accretion disk.

There is no significant difference between X-ray slope $\Gamma_X$ for
the different AGN types: mean $\Gamma_X=2.14 \pm 0.29$ for rapidly
accreting broad-line AGNs, and mean $\Gamma_X=2.05 \pm 0.29$ for
weakly accreting narrow-line and lineless AGNs (the difference is only
$1.2\sigma$ significant).  This is in contrast to the prediction of
\citet{hop09}, who suggest that harder X-ray slopes are expected for
radiatively inefficient accretion flows (RIAF) expected at low
accretion rates.  The appearance of a RIAF at inner radii might
produce more X-ray emission, as we discuss in \S 4.1 below, but this
emission probably has a similar power-law slope as the X-ray corona
present in broad-line AGNs with high accretion rates.  This is
unsurprising, since both the RIAF and the corona are thought to be
ionized plasmas with X-ray emission from inverse Compton scattering
and/or bremsstrahlung.  We can conclude that the onset of a RIAF in
unobscured narrow-line and lineless AGNs with accretion rates of
$10^{-4} < L_{int}/L_{Edd} < 10^{-2}$ do not cause harder X-ray
power-law slopes.

\subsection{Physics of the Accretion Disk}

As accretion rate increases from lineless and narrow-line to
broad-line AGNs, the disk temperature significantly increases and its
brightness with respect to the X-rays marginally increases.  An
increase in temperature with accretion rate is expected for a thin
accretion disk, which has $T_{max} \propto \dot{m}^{1/4}$
\citep{sha73}.  We additionally discuss below how the onset of a
radiatively inefficient accretion flow could also cause apparent
cooler disk emission.  Both an increase in temperature and in
$L_{disk}/L_X$ with accretion rates would contribute to the observed
increase of $\alpha_{OX}$ (the ratio of rest-frame UV to X-ray
emission) with accretion rate \citep{kel08,you10}.  In our previous
work \citep{tru09c}, we suggested that the increase of $\alpha_{OX}$
with accretion rate was due only to the disk luminosity decreasing
with respect to the corona luminosity.  While this is partly correct,
the correlation is also caused by increasing disk temperatures at
higher accretion rates.

AGNs with $L_{int}/L_{Edd} \lesssim 0.01$ are predicted to have
radiatively inefficient accretion flows (RIAFs) near the central black
hole \citep{beg84,nar95,yuan07,nar08}.  At such accretion rates, we
can define a truncation radius $R_t$ where the collisional cooling
time is comparable to the accretion time.  Beyond $R_t$, accretion
will remain in a standard geometrically thin and optically thick disk
with a thermal blackbody spectrum \citep[e.g.,][]{sha73}.  However
within $R_t$, there are too few collisions to couple the ions and
electrons and the gas becomes a two-temperature plasma.  The electrons
are cooled by bremsstrahlung, synchrotron, and Compton up-scattering,
while the ions remain at the virial temperature.  This means the flow
is geometrically thick and optically thin.  The introduction of a
truncation radius changes the $R_{in} = 6R_g$ assumption for the
accretion disk model, since by definition $R_{in} \geq R_t$.  The peak
energy of the best-fit accretion disk model is not very sensitive to
the choice of $R_{in}$, although larger inner radii change the shape
of the model with additional red emission.  At accretion rates
$L_{int}/L_{Edd} \gtrsim 10^{-3}$, as in our sample, $R_t \sim 80R_g$
\citep{yuan04}.  Using $r_{in}=80$ in the accretion disk model fitting
in Section 3.1 doesn't change the best-fit values of $E_{peak}$,
although it does result in slightly better fits.

The marginal ($2.1\sigma$ significant) increase of $L_{disk}/L_X$ with
$L_{int}/L_{Edd}$ might also be caused by the onset of the RIAF.  As
$R_t$ expands outwards, the disk emission decreases and the RIAF
emission increases.  The RIAF hot plasma emission is mostly X-ray
bremsstrahlung and Compton up-scattering (like the corona), with an
additional IR synchrotron component (which we discuss in Sections
4.3).  As accretion rate drops and $R_t$ increases, the rise of the
RIAF X-ray emission compared to the optical/UV disk emission is seen
as a decrease of $L_{disk}/L_X$.  Indeed, local low-luminosity AGNs
have even lower accretion rates and larger $R_t$, with consequently
lower $L_{disk}/L_X$ ratios and cooler optical thin-disk emission
\citep{ho08}.

The transition to an inner RIAF also causes the disappearance of broad
emission lines at $L_{int}/L_{Edd} \lesssim 0.01$.  \citet{nic00} was
the first to elegantly show that the broad emission lines are only
present above a critical accretion rate.  However \citet{nic00}
assumed that the innermost possible orbit was given by the
\citet{sha73} thin-disk model, $r_{crit} \simeq 8.16R_g$.  Here we
follow their basic derivation, with the key difference that we use the
RIAF transition radius as the innermost orbit for the presence of a
broad-line region.

There is evidence that the broad emission line region is part of a
disk wind \citep[e.g.][]{emm92,mur98,elv00,eli06}.  The positions of
individual broad emission lines are stratified and set by the ionizing
luminosity of the continuum \citep[e.g.][]{pet06,den09}.  The base of
the wind itself, however, is set by the radius at which the radiation
pressure equals the gas pressure, defined by \citet{sha73} as:
\begin{equation}
  \frac{r_{wind}}{(1-r_{wind}^{-0.5})^{16/21}}  \simeq 15.2 (\alpha M)^{2/21} \left(\frac{\dot{m}}{\eta}\right)^{16/21},
\end{equation}
with $r_{wind}$ in units of $R/(6R_g)=R/(6GM/c^2)$, $M$ in units of
$M_{BH}/M_{\odot}$, $\alpha$ is the viscosity parameter, and $\eta$ is
the accretion efficiency.  While RIAFs are expected to have strong
outflows (see Section 4.2), the RIAF region is a high-temperature
ionized plasma and so any associated disk wind would not emit broad
emission lines in the UV/optical.  Thus the RIAF truncation radius
sets the innermost possible radius for the existence of a broad-line
region.  Assuming that $\dot{m} \simeq L_{int}/L_{Edd}$ and
rearranging Equation 8 with $r_{wind} > R_t$, $\alpha \simeq 0.1$, and
$\eta \simeq 0.1$, this sets the minimum specific accretion rate for a
broad line region as:
\begin{equation}
  \dot{m} \gtrsim 0.013 (R_t/80R_g) M_8^{-1/8},
\end{equation}
with $M_8 = M_{BH}/(10^8 M_{\odot})$.  We leave $R_t$ as a free
parameter since it is poorly constrained, although the best-fit RIAF
models for $L_{int}/L_{Edd} \sim 10^{-3}-10^{-2}$ AGNs suggest $R_t
\sim 80R_g$ \citep{yuan04}.  As an AGN drops below this minimum
accretion rate, its broad lines disappear and only narrow lines (or no
lines) are observed, as seen in the transition at
$\log(L_{int}/L_{Edd}) \sim -2$ transition in Figures
\ref{fig:acchistogram} and \ref{fig:accretion}.

\citet{eli09} also predict that the disk wind associated with the BLR
will disappear below an accretion rate at which the outflowing
velocity drops below the random velocity of the disk.  \citet{eli09}
measure a BLR-disappearance accretion rate of $\log(L/L_{Edd}) < C +
\beta \log(L_{bol})$ from the low-luminosity local AGNs of
\citet{ho09}, with $\beta=-0.5$ and $C=14.4$.  In our sample \citep[as
well as those of][]{kol06,tru09b}, the BLR disappears at
$\log(L/L_{Edd}) < 0.01$.  For a typical bolometric luminosity of
$L_{int} \sim 10^{44.5}$~erg~s$^{-1}$ \citep[also appropriate for
the][sample]{kol06}, and assuming the same $\beta = -0.5$, this
instead corresponds to $C=20.3$: a remarkable difference of 6 orders
of magnitude.  It is unlikely that the bolometric corrections of
\citet{ho09} are incorrect by 6 orders of magnitude, and so we must
conclude that the \citet{eli09} model does not describe the
disappearance of the BLR for high luminosity AGNs.  Instead a
disk-wind model following \cite{nic00} best describes the BLR
disappearance as the radius of wind generation region moves within the
inner RIAF region.

It must be noted that while disk wind models have had success in
describing highly ionized emission and absorption lines in the UV
\citep{pro00,pro04}, they have not been applied to optical emission
lines.  The \Ha~broad emission line almost certainly forms in a higher
density, lower ionization region than the \CIV~and \MgII~broad
emission lines.  In addition there is evidence that the dynamics of
the \Hb~broad emission line are wildly variable, with reverberation
mapping indicating infalling, virialized, and outflowing \Hb~emission
regions in three AGNs \citep{den09}.  While we do find that broad
\Hb~tends to be present only for $L_{int}/L_{Edd} \gtrsim 0.01$ and so
fits in the wind/RIAF framework, we do not study \Ha~and cannot say if
this line is described by the same physics.  Indeed, \citet{ho09}
present several AGNs with broad \Ha~emission and $L_{int}/L_{Edd} <
10^{-3}$.  This suggests that broad \Ha~emission may have its origin
outside the disk wind, although it is important to note that the
accretion rates of \citet{ho09} rely on bolometric corrections to
monochromatic luminosities and so may suffer from significant
systematic uncertainties.

\subsection{Accretion Rate and Outflows}

\begin{figure}
\scalebox{1.2}
{\plotone{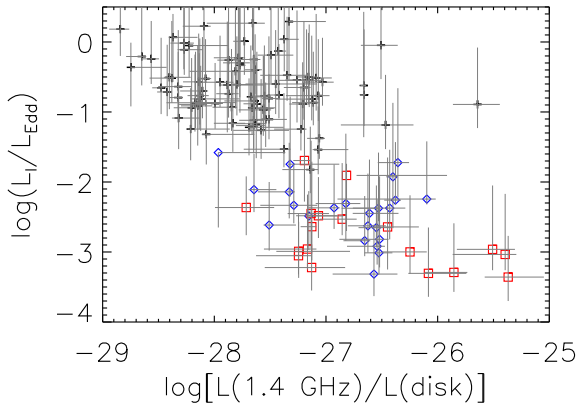}}
\figcaption{Accretion rate with a measure of radio brightness: the
  ratio of radio luminosity to disk luminosity for the 118 unobscured
  ($N_H<10^{22}$~cm$^{-2}$) AGNs in our sample.  Broad-line AGNs are
  shown by black crosses, narrow-line AGNs by blue diamonds, and
  lineless AGNs by red squares.  Narrow-line and lineless AGNs, at
  lower accretion rates than broad-line AGNs, tend to be more radio
  luminous compared to their accretion disk luminosity.
\label{fig:radio}}
\end{figure}

The gas in a RIAF is not gravitationally bound to the supermassive
black hole because the ions are not losing energy through radiation.
As a result, AGNs with RIAFs are predicted to have strong radio
outflows \citep{nar95,mei01}.  The coupling between a RIAF and a
strong radio outflow has been confirmed by observations of black hole
binaries \citep{fen04}, and it is possible to translate these
observations to AGN scales \citep[e.g.][]{mac03}.  In Figure
\ref{fig:radio} we show the AGNs of our sample with the ratio of radio
luminosity to disk luminosity.  Note that since the radio emission is
coincident with the X-ray point source we assume that it originates
from the AGN, but we cannot strictly rule out other sources of radio
emission (e.g., from star formation).  The $L_{int}/L_{Edd}<10^{-2}$
AGNs which are expected to have RIAFs tend to have higher ratios of
radio to disk (optical/UV) luminosity.  The mean $L_{disk}/L_{radio}$
for rapidly accreting broad-line AGNs is a factor of ten lower than
the mean $L_{disk}/L_{radio}$ for narrow-line and lineless AGNs, and
since the scatter in each sample is about $\sim$0.5~dex this
translates to a highly significant difference ($14.9\sigma$).

The large scatter in the $L_{disk}/L_{radio}$ ratio at both high and
low accretion rates is likely because the radio power is additionally
dependent on properties like black hole spin and orientation.  But the
highly significant increase in $L_{disk}/L_{radio}$ for low accretion
rate AGNs suggests that $L_{int}/L_{Edd}<10^{-2}$ AGNs with RIAFs
generally have relatively brighter radio emission.  \citet{mel10}
noticed a similar trend of increasing radio luminosity with decreasing
accretion rate, using \OIV as a proxy for intrinsic luminosity
\citep[e.g.,][]{mel08,dia09}.  Many nearby radio galaxies are also
measured to have low accretion rates and may even have their
optical/UV emission dominated by synchrotron emission rather than a
thermal disk \citep{chi99}.

In general, the radiation and disk winds of AGNs are thought to cause
feedback on galaxy scales by quenching star formation
\citep[e.g.,][]{hop06,hop10}, while radio jets are thought to cause
larger-scale feedback which can heat the cores of galaxy clusters
\citep[e.g.,][]{fab02} and is observed as extended emission line
regions \citep{fu09}.  The fact that RIAFs tend to have stronger radio
emission suggests that weakly accreting AGNs may remain important for
large-scale radio-mode feedback despite their optical/UV and X-ray
luminosities.  This suggests that heating cluster cores may not
require bright quasars, but can be accomplished by faint AGNs
\citep[see also][]{har09}.  \citet{all06} similarly found that several
nearby weakly accreting AGNs had most of their Bondi accretion rates
converted to radio outflows.

\subsection{Accretion Rate and the IR ``Torus''}

\begin{figure}
\scalebox{1.2}
{\plotone{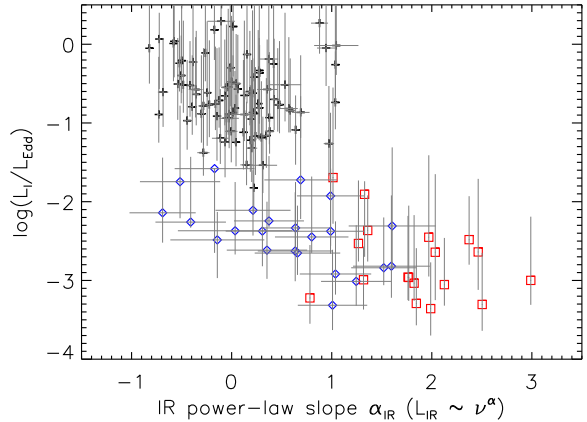}}
\figcaption{Accretion rate with the power-law slope of the
  $1<\lambda<10\mu$m IR emission for the 118 unobscured
  ($N_H<10^{22}$~cm$^{-2}$) AGNs.  As in previous figures, black
  crosses are broad-line AGNs, blue diamonds are narrow-line AGNs, and
  red squares are lineless AGNs.  We measure the slope $\alpha_{IR}$
  as $L \sim \nu^{\alpha}$, corresponding to the slope $\beta$ used in
  the power-law selection of \citet{don07} as $\beta=\alpha_{IR}-1$.
  Most high accretion rate ($L_{int}/L_{Edd}>0.01$) AGNs have IR
  power-law slopes corresponding to a dusty torus ($\alpha_{IR}<0.5$).
  Of $L_{int}/L_{Edd}<0.01$ AGNs, however, half the narrow-line and
  all but one of the lineless AGNs lack the torus signature.
\label{fig:iraccrete}}
\end{figure}

A clumpy dust ``torus'' emits a unique power-law signature in the
mid-IR from $\sim$1-10$\mu$m \citep{nen08}.  This was first noticed
observationally as a distinct AGN locus in Spitzer/IRAC color-color
space \citep{lacy04,stern05}, although \citet{don07} show that
power-law selection is the most effective way to select AGN in the
mid-IR.  We compute the IR power-law slope in our AGNs from the
host-subtracted observed IRAC photometry within the rest-frame
wavelength range $1<\lambda<10\mu$m, shown with accretion rate in
Figure \ref{fig:iraccrete}.  Type 1 AGNs typically have
$\alpha_{IR}<0.5$ \citep[$\beta_{IR}<-0.5$ in terms of the $f_{\nu}
\sim \nu^{\beta}$ form used by][]{don07}, matching the predictions of
clumpy dust models \citep{nen08}.  About 10\% of Type 1 AGNs are
``hot-dust-poor'' and do not satisfy the $\alpha_{IR}$ selection
criterion\footnote{For more details on this population, see
\citet{hao10}.}, about half of the narrow-line AGNs and only one of
the lineless AGNs lack this torus signature.  \citet{car08} similarly
found that many X-ray AGNs did not have a mid-IR power-law, although
they did not track it with accretion rate.  In our sample, the rapidly
accreting broad-line AGNs have a mean $\alpha_{IR}=0.38 \pm 0.35$,
while the weakly accreting narrow-line and lineless AGNs have a mean
$\alpha_{IR}=1.26 \pm 0.72$, meaning that the two samples differ with
high significance ($10.7\sigma$).

A unified model based solely on geometrical obscuration suggests that
narrow-line and lineless AGNs are obscured by the same torus present
in broad-line AGNs \citep[e.g.,][]{ant93}.  Instead the low accretion
rate AGNs ($L_{int}/L_{Edd}<0.01$) frequently lack the torus IR
signature.  In part, this may be because the torus power-law is simply
being overwhelmed by the accretion disk SED at $L_{int}/L_{Edd}<0.01$.
At low accretion rates, the temperature of the disk decreases, and a
disk with $E_{peak}=1$~eV will peak at 1.2~$\mu$m, emitting a
power-law of $\alpha \sim 2$ at $1<\lambda<10 \mu$m.  In a typical
broad-line AGN, the IR torus is roughly the same strength as the
accretion disk \citep[][see also Figure \ref{fig:sedfits}]{ric06}.
Since many $L_{int}/L_{Edd}<0.01$ AGNs in Figure \ref{fig:iraccrete}
have $\alpha \gtrsim 2$, they must be dominated by the accretion disk
emission and have, at best, very little emission from the torus.

The weaker or missing torus in many $L_{int}/L_{Edd}<0.01$ AGNs can be
described in a similar fashion to the vanishing disk-wind BLR in
Section 4.1.  There is good evidence that the outer edge of the BLR
coincides with the inner edge of the clumpy dust \citep{net93,sug06}.
Some authors additionally suggest that the BLR and the clumpy dust
``torus'' are two components of the same wind driven off the accretion
disk \citep[e.g.,][]{eli06}.  If the clumpy dust wind emerges from the
disk at a similar radius to that calculated in Section 4.1, then we
would expect the IR power-law signature to disappear at
$L_{int}/L_{Edd}<0.01$, just as the BLR disappears.  However many
narrow-line AGNs with $L_{int}/L_{Edd}<0.01$ still have the negative
IR power-law slopes, suggesting that there must be another source of
mid-IR emission.  Either there is a distant source of clumpy dust
beyond the expanding RIAF, or there is mid-IR synchrotron emission in
the RIAF region at the base of the radio jet \citep[as observed
  by][]{lei09}.

\section{A Simple Model for Unifying AGNs by Specific Accretion Rate}

\begin{figure*}
\centering
\scalebox{1.1}
{\plotone{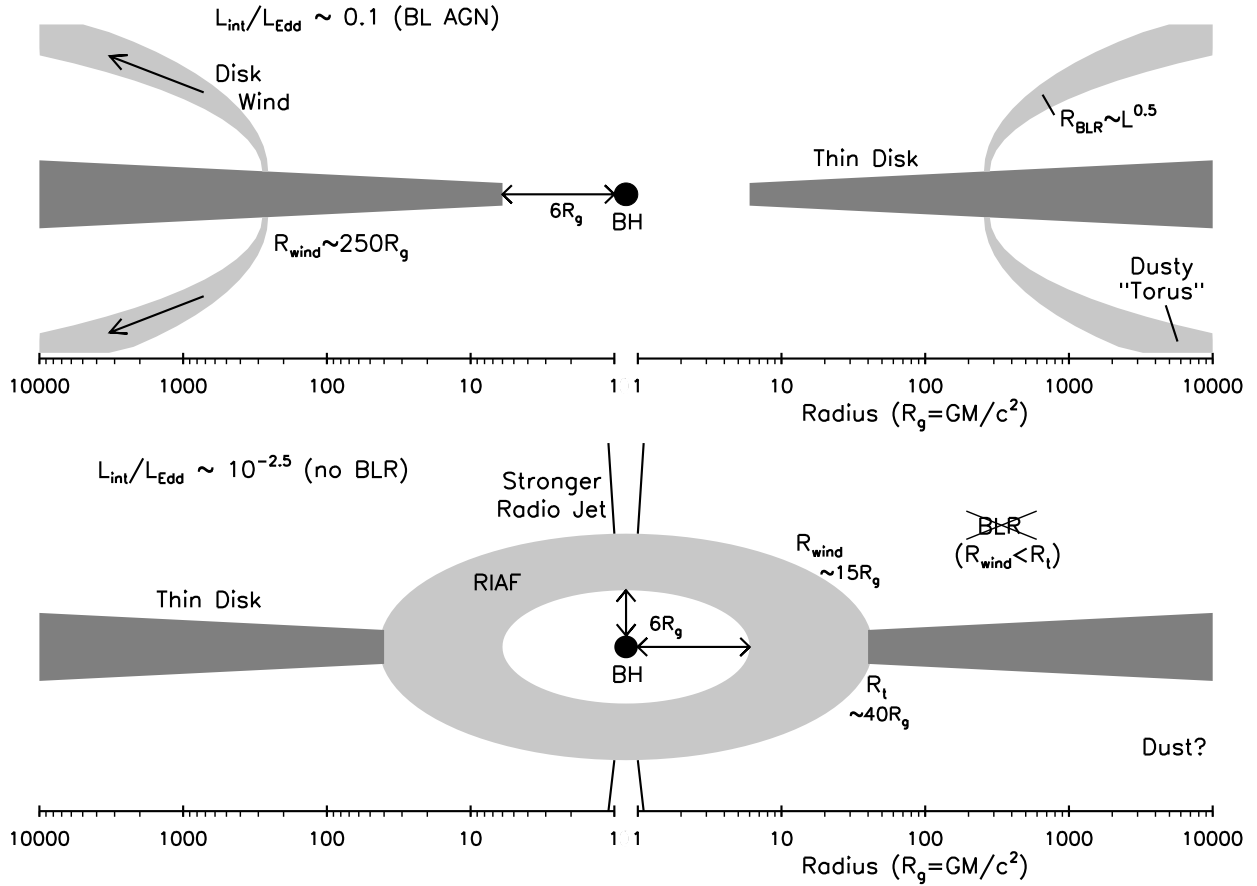}}
\figcaption{A schematic model showing the changes in the accretion
  disk from a broad-line AGN with high accretion rate
  ($L_{int}/L_{Edd} \sim 0.1$) to a narrow-line or lineless AGN with
  low accretion rate ($L_{int}/L_{Edd} \sim 0.003$).  The x axis shows
  the radial distance from the black hole in units of $GM/c^2$.  The y
  axis is qualitative only.  At $L_{int}/L_{Edd} \lesssim 0.01$, the
  disk wind falls inside the RIAF.  As a result there are no broad
  emission lines, the hot dust signature becomes very different, and
  the radio jet becomes stronger.
\label{fig:cartoonmodel}}
\end{figure*}

Figure \ref{fig:cartoonmodel} presents a simple schematic outlining
the changes in AGNs from high ($L_{int}/L_{Edd}>0.01$) to low
($L_{int}/L_{Edd}<0.01$) accretion rate.  At the top is a broad-line
AGN with high accretion rate ($L_{int}/L_{Edd} \sim 0.1$).  At these
high accretion rates the gas and dust falling into the black hole
forms a thin accretion disk and a disk wind originates at $R_{wind}
\sim 250R_g$.  The broad emission lines are emitted in stratified
regions along this wind based on the radiation pressure (which ionizes
and excites the wind material), with $R_{BLR} \sim L^{0.5}$ and high
ionization lines (e.g., \CIV) emitted from nearer radii than low
ionization lines (e.g., \MgII) \citep{pet06}.  At higher radii, the
disk wind forms clumpy dust \citep{nen08}.  This dusty ``torus'' can
obscure the AGN along lines of sight near the disk, causing an
observer to see an obscured narrow-line AGN \citep{ant93}.

The bottom of Figure \ref{fig:cartoonmodel} shows an AGN with low
accretion rate ($L_{int}/L_{Edd} \sim 0.003$), characteristic of the
unobscured narrow-line and lineless AGNs in our sample.  The onset of
a geometrically thick RIAF changes the picture dramatically.  Because
the disk wind radius is within the RIAF, there are no broad emission
lines.  Instead the dominant outflow is a radio jet, and AGNs with low
accretion rates and RIAFs are typically more radio luminous than
broad-line AGNs.  The lack of a disk wind also means that there is not
the typical clumpy dust ``torus'' seen in broad-line AGNs.  However we
cannot rule out the presence of dust completely, as clumpy dust may
come from another source besides the disk wind and some
$L_{int}/L_{Edd} \lesssim 0.01$ have the IR signature of hot dust.

\section{Predictions and Future Observational Tests}

The multiwavelength data of COSMOS provide many diagnostic
capabilities, and we have argued that decreasing accretion rates lead
to the onset of a RIAF at $\dot{m}<0.01$ and subsequently stronger
radio jets, a weaker torus, and the disappearance of broad emission
lines.  The onset of a RIAF also makes several predictions testable by
future observations.  In addition the simple model in Section 5 can be
more fully constrained by additional investigations.

If the broad-line region is truly disappearing at $\dot{m}<0.01$ then
we would expect spectropolarimetry to reveal reflected broad emission
lines in only high accretion rate ($\dot{m}>0.01$) narrow-line and
lineless AGNs.  Spectropolarimetry of nearby AGNs shows a dichotomy
based on accretion rate, although most authors place the change from
hidden broad lines to ``true'' Type 2 AGN at $\dot{m} \sim 0.001$
\citep{tran03,wang07}.  Most likely, the difference results from the
uncertain bolometric corrections used in these previous works,
compared to the full modeled SEDs used here.

Mid-IR broad-band polarimetry could determine the cause of the
negative IR power-law slopes in $\dot{m}<0.01$ AGNs.  If the clumpy
dust ``torus'' is associated with the same wind that drives the broad
line region, it should disappear in these objects.  The mid-IR
signature might instead be synchrotron radiation in the RIAF at the
base of the jet, which would appear polarized at the $>3\%$ level
\citep[e.g.,][]{jan94}.  If no polarization is detected, then we must
conclude that clumpy dust exists at higher radii than the BLR
disk-wind, beyond the RIAF region of $\dot{m}<0.01$ AGNs.

It is very difficult to measure accretion rates of partially or fully
obscured AGNs, and such objects are generally missed by the X-ray and
optical limits of this study.  However we do make a few predictions
for the accretion rates of various AGNs.  If the torus is part of a
disk-wind that vanishes at $\dot{m}<0.01$, then torus-obscured AGNs of
the classical \citet{ant93} unified model will have only high
accretion rates ($\dot{m}>0.01$).  Obscuration by cooler dust
associated with host galaxy star formation, as predicted by the
observed redshift evolution in the narrow-line/broad-line AGN ratio
\citep{trei09,tru09a}, could conceivably be present at any accretion
rate \citep[although it may be limited by the ability of the dusty
  star formation to feed the black hole,][]{bal08}.  We might then
expect that obscured AGNs with a strong mid-IR torus signature should
have $\dot{m}>0.01$, while AGNs obscured by the cooler dust associated
with host galaxy star formation might have a wider range of accretion
rates.

\acknowledgements

JRT acknowledges support from NSF/DDEP grant \#0943995, and with CDI
acknowledges support from NSF grant \#AST-0908044.  BCK acknowledges
support from NASA through Hubble Fellowship grant \#HF-51243.01
awarded by the Space Telescope Science Institute, which is operated by
the Association of Universities for Research in Astronomy, Inc., for
NASA, under contract NAS 5-26555.

\begin{deluxetable}{lcrcrrr}
\tablecolumns{7}
\tablewidth{0pc}
\tablecaption{Catalog of AGNs\tablenotemark{a}
\label{tbl:catalog}}
\tablehead{
  \colhead{RA+Dec (J2000)} &
  \colhead{Type\tablenotemark{b}} & 
  \colhead{Redshift} & 
  \colhead{Spec.\tablenotemark{c}} & 
  \colhead{$L_{int}$} & 
  \colhead{$M_{BH}$} &
  \colhead{$\log(L_{int}/L_{Edd})$} \\
  \colhead{hhmmss.ss+ddmmss.s} &
  \colhead{} &
  \colhead{} &
  \colhead{source} &
  \colhead{$\log(L_{\odot})$} &
  \colhead{$\log(M_{\odot})$} &
  \colhead{} }
\startdata
095728.34+022542.2 & BL & 1.54 & S & $46.03_{-0.10}^{+ 0.52}$ & $ 8.40_{-0.43}^{+ 0.36}$ & $-0.49_{-0.27}^{+ 0.58}$ \\
095740.78+020207.9 & BL & 1.48 & I & $45.88_{-0.30}^{+ 0.64}$ & $ 8.24_{-0.39}^{+ 0.45}$ & $-0.47_{-0.40}^{+ 0.68}$ \\
095743.33+024823.8 & BL & 1.36 & S & $45.84_{-0.16}^{+ 0.66}$ & $ 8.24_{-0.36}^{+ 0.44}$ & $-0.51_{-0.66}^{+ 0.18}$ \\
095749.02+015310.1 & NL & 0.32 & I & $43.89_{-0.21}^{+ 0.71}$ & $ 8.61_{-0.30}^{+ 0.29}$ & $-2.84_{-0.15}^{+ 0.73}$ \\
095750.20+022548.3 & BL & 1.24 & Z & $44.93_{-0.20}^{+ 0.52}$ & $ 7.28_{-0.41}^{+ 0.38}$ & $-0.46_{-0.28}^{+ 0.60}$ \\
095752.17+015120.1 & BL & 4.16 & Z & $46.28_{-0.10}^{+ 0.69}$ & $ 8.71_{-0.42}^{+ 0.41}$ & $-0.54_{-0.36}^{+ 0.53}$ \\
095752.17+015120.1 & BL & 4.17 & I & $46.26_{-0.07}^{+ 0.54}$ & $ 8.66_{-0.44}^{+ 0.38}$ & $-0.51_{-0.31}^{+ 0.52}$ \\
095753.49+024736.1 & BL & 3.61 & I & $46.24_{-0.27}^{+ 0.75}$ & $ 8.00_{-0.40}^{+ 0.49}$ & $ 0.12_{-0.46}^{+ 0.69}$ \\
095754.11+025508.4 & BL & 1.57 & S & $46.21_{-0.49}^{+ 0.66}$ & $ 8.70_{-0.41}^{+ 0.39}$ & $-0.61_{-0.32}^{+ 0.65}$ \\
095754.70+023832.9 & BL & 1.60 & S & $46.14_{-0.24}^{+ 0.54}$ & $ 8.72_{-0.41}^{+ 0.40}$ & $-0.69_{-0.39}^{+ 0.47}$ \\
095755.08+024806.6 & BL & 1.11 & S & $45.94_{-0.17}^{+ 0.62}$ & $ 8.43_{-0.43}^{+ 0.36}$ & $-0.60_{-0.28}^{+ 0.57}$ \\
095755.34+022510.9 & BL & 2.74 & I & $45.51_{-0.07}^{+ 0.45}$ & $ 8.07_{-0.45}^{+ 0.36}$ & $-0.68_{-0.36}^{+ 0.44}$ \\
095755.48+022401.1 & BL & 3.10 & I & $46.82_{-0.47}^{+ 0.78}$ & $ 8.44_{-0.45}^{+ 0.42}$ & $ 0.27_{-0.24}^{+ 0.79}$ \\
095759.50+020436.1 & BL & 2.03 & S & $46.82_{-0.46}^{+ 0.70}$ & $ 8.94_{-0.40}^{+ 0.38}$ & $-0.23_{-0.56}^{+ 0.32}$ \\
095759.91+021634.5 & BL & 1.54 & I & $44.68_{-0.10}^{+ 0.67}$ & $ 8.26_{-0.39}^{+ 0.45}$ & $-1.70_{-0.37}^{+ 0.46}$ \\
095801.61+020428.9 & BL & 1.23 & Z & $45.43_{-0.26}^{+ 0.76}$ & $ 8.28_{-0.39}^{+ 0.33}$ & $-0.96_{-0.26}^{+ 0.76}$ \\
095802.10+021541.0 & OD & 0.94 & I & $45.05_{-0.06}^{+ 0.59}$ & $ 9.44_{-0.33}^{+ 0.38}$ & $-2.50_{-0.33}^{+ 0.47}$ \\
095805.10+020445.8 & NL & 0.67 & I & $44.91_{-0.08}^{+ 0.55}$ & $ 9.33_{-0.39}^{+ 0.34}$ & $-2.54_{-0.20}^{+ 0.72}$ \\
095806.24+020113.8 & NL & 0.62 & I & $44.25_{-0.28}^{+ 0.65}$ & $ 9.03_{-0.33}^{+ 0.35}$ & $-2.90_{-0.34}^{+ 0.70}$ \\
095806.99+022248.5 & BL & 3.10 & I & $46.39_{-0.25}^{+ 0.69}$ & $ 9.34_{-0.43}^{+ 0.36}$ & $-1.06_{-0.31}^{+ 0.68}$ \\
095809.45+020532.4 & OD & 0.61 & I & $44.00_{-0.33}^{+ 0.48}$ & $ 9.26_{-0.29}^{+ 0.43}$ & $-3.38_{-0.43}^{+ 0.52}$ \\
\enddata
\tablenotetext{a}{The full catalog will appear as a machine-readable
  table in the electronic version.}
\tablenotetext{b}{``BL'' refers to a broad-line AGN, ``NL'' is a
  narrow-line AGN, and ``OD'' is a lineless or optically dull AGN.}
\tablenotetext{c}{``S'' means the spectrum and redshift are from the
  SDSS archive, ``I'' is from the COSMOS Magellan/IMACS campaign
  \citep{tru09a}, and ``Z'' is from the zCOSMOS VLT/VIMOS campaign
  \citep{lil10}.}
\end{deluxetable}

\end{document}